\documentclass[twocolumn,showpacs,aps,floatfix,prd,nofootinbib]{revtex4}

\usepackage{graphicx}
\usepackage{dcolumn}
\usepackage{array, longtable}
\usepackage{epsfig}
\usepackage{amsmath}
\usepackage{colordvi}
\usepackage{hhline}

\newcommand{\BaBarYear}{16}
\newcommand{\BaBarNumber}{004}
\newcommand{\SLACPubNumber}{16918}

 \newcommand{\BaBarType}      {PUB}  

\input babarsym

\def\Ecm       {\ensuremath {E_{\rm c.m.}}\xspace}

\def\KKpz      {\ensuremath {\KS\KL\piz}\xspace}
\def\KKeta     {\ensuremath {\KS\KL\eta}\xspace}
\def\KKppz     {\ensuremath {\KS\KL\piz\piz}\xspace}

\def\chiKKg    {\ensuremath {\chi^2(\KS\KL)}\xspace}
\def\chiKKpg   {\ensuremath {\chi^2(\KS\KL\piz)}\xspace}
\def\chiKKeg   {\ensuremath {\chi^2(\KS\KL\eta)}\xspace}
\def\chiKKppg  {\ensuremath {\chi^2(\KS\KL\piz\piz)}\xspace}

\def\jetset   {\mbox{\tt Jetset \hspace{-0.5em}7.\hspace{-0.2em}4}\xspace}

\long\def\inst#1{\par\nobreak\kern 4pt\nobreak
    {\it #1}\par\vskip 10pt plus 3pt minus 3pt}

\begin{document}

\begin{flushleft}
\babar-\BaBarType-\BaBarYear/\BaBarNumber \\
SLAC-PUB-\SLACPubNumber \\
\end{flushleft}

\title{\large \bf
\boldmath
Cross sections for the reactions 
$\epem\to \KKpz$, \KKeta, and \KKppz    
from events with initial-state radiation
} 

%
\author{J.~P.~Lees}
\author{V.~Poireau}
\author{V.~Tisserand}
\affiliation{Laboratoire d'Annecy-le-Vieux de Physique des Particules (LAPP), Universit\'e de Savoie, CNRS/IN2P3,  F-74941 Annecy-Le-Vieux, France}
\author{E.~Grauges}
\affiliation{Universitat de Barcelona, Facultat de Fisica, Departament ECM, E-08028 Barcelona, Spain }
\author{A.~Palano}
\affiliation{INFN Sezione di Bari and Dipartimento di Fisica, Universit\`a di Bari, I-70126 Bari, Italy }
\author{G.~Eigen}
\affiliation{University of Bergen, Institute of Physics, N-5007 Bergen, Norway }
\author{D.~N.~Brown}
\author{Yu.~G.~Kolomensky}
\affiliation{Lawrence Berkeley National Laboratory and University of California, Berkeley, California 94720, USA }
\author{M.~Fritsch}
\author{H.~Koch}
\author{T.~Schroeder}
\affiliation{Ruhr Universit\"at Bochum, Institut f\"ur Experimentalphysik 1, D-44780 Bochum, Germany }
% ...
\collaboration{The \babar\ Collaboration}
\noaffiliation


\begin{abstract}
We study the processes $\epem\to\KKpz\gamma$, $\KKeta\gamma$, and
$\KKppz\gamma$, 
where the photon is radiated from the initial state,
providing cross section measurements for the hadronic final states
over a continuum of center-of-mass energies.
The results are based on 469~\invfb of
  data collected 
at or near the \FourS~ resonance
with the \babar\ detector at SLAC.  
We present the first measurements of the 
 $\epem\to\KKpz$, \KKeta, and $\KKppz$ cross sections up to a
 center-of-mass energy of 4~\gev, 
and study their intermediate resonance structures.  
We observe \jpsi decays to all of these final states for the first time,
present measurements of their \jpsi branching fractions, 
and search for \psitwos decays.

\end{abstract}


\vfill
\maketitle

\setcounter{footnote}{0}

\section{Introduction}
\label{sec:Introduction}

Electron-positron annihilation events with initial-state radiation
(ISR) can be used to study processes over a wide range of energies
below the nominal \epem center-of-mass (c.m.) energy (\Ecm),
as demonstrated in Ref.~\cite{baier}.
The possibility of exploiting ISR to make precise measurements of
low-energy cross sections at high-luminosity $\phi$ and $B$ factories
is discussed in Refs.~\cite{arbus, kuehn, ivanch},
and motivates the studies described in this paper.  
Such measurements are of particular current interest because of a
three-standard-deviation discrepancy between the measured value of the
muon anomalous magnetic moment ($g_\mu-2$) and that computed 
in the Standard Model~\cite{dehz},  
where the hadronic loop contributions require experimental 
\epem annihilation cross sections as input.
The calculation is most sensitive to the low-energy region, 
where the inclusive hadronic cross section cannot be measured
reliably, 
and a sum of exclusive states must be used.
Not all accessible states have been measured yet, 
and new measurements will improve the reliability of the calculation.
In addition, studies of ISR events at $B$ factories provide
information on resonance spectroscopy for masses up through the
charmonium region. 
                                
Studies of the ISR processes $e^+e^-\to\mumu\gamma$~\cite{Druzhinin1,isr2pi}
and $\epem \to X_h\gamma$, 
where $X_h$ represents any of several exclusive multihadron final states,
using data of the \babar\ experiment at SLAC,
have been reported previously.
The $X_h$ studied so far include:
charged hadron pairs $\pip\pim$~\cite{isr2pi}, $\Kp\Km$~\cite{isr2k}, and
$p\overline{p}$~\cite{isr2p};
four or six charged mesons~\cite{isr4pi,isr2k2pi,isr6pi};
charged mesons plus one or two \piz mesons~\cite{isr2k2pi,isr6pi,isr3pi,isr5pi};
a \KS plus charged and neutral mesons~\cite{isrkkpi}; 
and the first ISR measurement from \babar\ that includes \KL
mesons~\cite{isrkskl}. 
Together, they demonstrate good detector efficiency for events of
this kind, 
and well understood tracking, particle identification, and \piz,
\KS and \KL reconstruction.

In this paper we report measurements of the \KKpz, \KKeta, and \KKppz final
states,
produced in conjunction with a hard photon that is assumed to result
from ISR. 
Candidate \KS mesons are reconstructed in the $\pip\pim$ decay mode,
candidate \piz and $\eta$  mesons are reconstructed in the
$\gamma\gamma$ decay mode, 
and \KL mesons are detected via their nuclear interactions in the
electromagnetic calorimeter.
For these final states, 
we measure cross sections from threshold to $\Ecm = 4$~\gev, 
study their internal structure,
perform the first measurements of \jpsi branching fractions,
and search for \psitwos decays.
We also search for the $\epem\to\gamma\KS \KS\piz$ and 
$\epem\to\gamma\KS\KS\piz\piz$ processes, 
which are forbidden by $C$-parity conservation,
and we see no indication of them 
at the level of single background events.
Together with our previous 
measurements~\cite{isr2k,isr2k2pi,isrkskl}, 
these results provide a much more complete understanding of the 
$K\Kbar\pi$, $K\Kbar\eta$ and $K\Kbar\pi\pi$ final states in \epem
annihilation.

\section{\boldmath The \babar\ detector and data set}
\label{sec:babar}

The data used in this analysis were collected with the \babar\ detector at
the \pep2\ asymmetric-energy \epem\ storage ring. 
The total integrated luminosity used is 468.6~\invfb~\cite{lumi}, 
which includes data collected at the $\Upsilon(4S)$
resonance (424.7~\invfb) and at a c.m.\ energy 40~\mev below this
resonance (43.9~\invfb). 

The \babar\ detector is described in detail elsewhere~\cite{babar}. 
Charged particles are reconstructed using the \babar\ tracking system,
which comprises the silicon vertex tracker (SVT) and the drift chamber (DCH)
inside the 1.5 T solenoid.
Separation of pions and kaons is accomplished by means of the detector of
internally reflected Cherenkov light (DIRC) and energy-loss measurements in
the SVT and DCH. 
The hard ISR photon, photons from $\pi^0$ and $\eta$ decays,
and \KL are detected in the electromagnetic calorimeter (EMC).  
Muon identification, provided by the instrumented flux return,
is used to select the $\mumu\gamma$ final state.

To study the detector acceptance and efficiency, 
we have developed a special package of simulation programs for
radiative processes based on 
the approach suggested by K\"uhn  and Czy\.z~\cite{kuehn2}.  
Multiple collinear soft-photon emission from the initial \epem state 
is implemented with the structure-function technique~\cite{kuraev,strfun}, 
while additional photon radiation from the final-state particles is
simulated using the PHOTOS package~\cite{PHOTOS}.  
The precision of the radiative simulation contributes less than 1\% to
the uncertainty of the measured hadronic cross sections.

In addition to the signal channels \KKpz, \KKeta, and \KKppz, 
we simulate ISR processes which result in high backgrounds,
$\KS\KL$, $\KS\Kpm\pimp$, and $\KS\Kpm\pimp\piz$,
with cross sections and mass dependences based on our previous
measurements and isospin relations.
The $\KS\KL$ and \KKeta channels are dominated by $\epem\to\gamma\phi$
and $\gamma\phi\eta$, respectively.
Samples of three to five times the number of expected events are
generated for each final state and processed through the detector
response simulation \cite{GEANT4}.  
These events are then reconstructed using the same software chain
as the data. 

We also simulate several non-ISR backgrounds, including 
$\epem\to q \qbar$ $(q = u, d, s, c)$ events using the
\jetset~\cite{jetset} generator,
and $\epem\to\tau^+\tau^-$ events using the KORALB~\cite{koralb} generator. 
Variations in detector and background conditions are taken into account.
\begin{figure}[tbh]
\begin{center}
\includegraphics[width=0.9\linewidth]{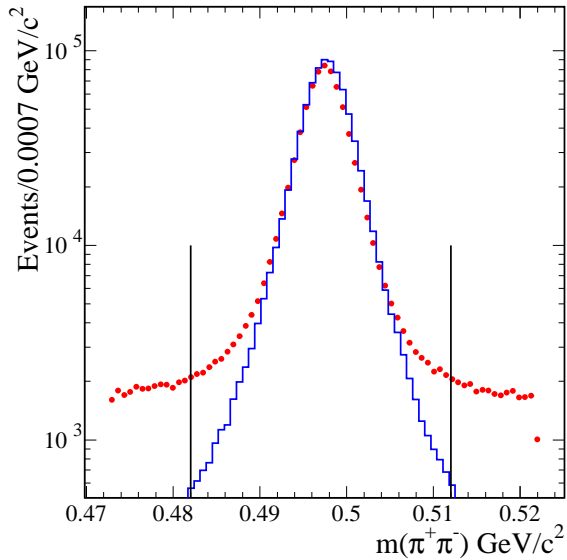}
\vspace{-0.4cm}
\caption{The $\pipi$ invariant mass distribution for the selected
  $\KS$ candidates in the data (points) and simulation (histogram).
  The vertical lines indicate the signal region.
}
\label{ksmass}
\end{center}
\end{figure}
\section{\boldmath Event Selection and Reconstruction}
\label{sec:KS}
We begin with events containing at least two charged particles and at
least four clusters of energy deposits in contiguous crystals in the EMC.
We then consider the cluster in the event with the highest energy in the
\epem c.m.\ frame as the ISR photon candidate, 
and require $E^\gamma_{\rm c.m.} > 3~\gev$.
Since the ISR photons are produced mostly along the beamline, 
this accepts only about 10\% of the signal events,
but in the selected events, 
the hadronic system is fully contained and can be studied reliably.

In these events, 
we reconstruct candidate \KS decays to two charged pions from pairs
of oppositely charged tracks not identified as electrons.
They must have a well reconstructed vertex between 0.1 and 40.0~cm in
radial distance from the beam axis,
and their total momentum must be 
consistent with the assumption that
        they originate from the interaction region.
The $m(\pipi)$ invariant mass distribution for these \KS candidates 
is shown in Fig.~\ref{ksmass} for both data (points) and the
$\epem \to \gamma\phi \to \gamma\KS\KL$ simulation (histogram).
The signal is very clean, and requiring $482< m(\pipi)<512~\mevcc$
(vertical lines in Fig.~\ref{ksmass}) accepts 98\% of the signal
events.
We use the sidebands 472--482 and 512--522~\mevcc to estimate the
contributions from non-\KS backgrounds, 
which are found to be negligible in all cases after final
selection.

A few thousand events (about 1\% of the total) have more than one
selected \KS candidate,  
and we use only the candidate with $m(\pipi)$ closest to the
nominal~\cite{PDG} \KS mass.
We also require the event to contain no other tracks that 
extrapolate within 2~cm of the beam axis and 3~cm along the axis from
the nominal interaction point.

Any number of additional tracks and EMC clusters is allowed.
We consider all clusters with reconstructed energy above 0.1~\gev as photon
candidates, and calculate the invariant mass of each pair.
Every pair with a mass within 30 (50)~\mevcc of the nominal \piz
($\eta$) mass is considered a \piz ($\eta$) candidate.
The efficiency of \piz and $\eta$ reconstruction in these events is
about 97\%.

The decay length of the \KL meson is large, 
and the probability to detect a \KL decay in the DCH is low. 
Instead, we look for a cluster in the EMC resulting from the
interaction of a \KL with a nucleus in the EMC material.
Such clusters are indistinguishable from photon-induced clusters, 
and give poor resolution on the \KL energy.
The characteristics of these clusters were studied in detail in our previous
publication~\cite{isrkskl}, 
where it was shown using $\epem\to\phi\gamma$ events that \KL clusters
are detected with high efficiency and good angular resolution.
Background from low-energy clusters is high, and the requirement of
at least 0.2~\gev in cluster energy yields a clean sample with 48\%
efficiency. 
Here, we apply the same energy requirement, and use the efficiency and
angular resolution measured as a function of polar and azimuthal angles
in Ref.~\cite{isrkskl}.
\section{\boldmath The kinematic fit procedure}
\label{sec:Analysis}
Each event selected as described in Sec.~\ref{sec:KS} is
subjected to a set of constrained kinematic fits,
in which the four-momenta and covariance matrices of 
the initial \epem, the ISR photon, the best \KS candidate,
and zero, two or four relevant photon candidates
are taken into account. 
The direction and angular resolution, but not the energy, of the \KL
candidate is also used,
and the \KL momentum is determined in the fit.
The three-momentum vectors for all other particles, including the photons, 
are also determined with better accuracy from the fits, and the fitted values
are used in further calculations. 

For every event, we first perform kinematic fits under the $\KS \KL \gamma$ hypothesis, 
considering the ISR photon and \KS candidates, along with each cluster
with energy over 0.2~\gev in turn as the \KL candidate.
Each fit has 3 constraints, and we consider the combination with the best
\chisq value, denoted \chiKKg. 
This variable is useful in suppressing the large background
arising from combinations of background photons with a mass near
the $\piz$ or $\eta$ mass. 

Next, we consider each \piz and $\eta$ candidate, 
and perform a set of fits to the $\KS\KL\piz\gamma$ and
$\KKeta\gamma$ hypothesis, including the ISR photon, \KS, and two
photon candidates, along with each \KL candidate not included in the
\piz or $\eta$ candidate.
These fits have four constraints, including one on the \piz or $\eta$
candidate mass.  
We retain the $\piz\KL$ and $\eta\KL$ candidate combinations
yielding the best values of \chiKKpg and \chiKKeg, respectively.

Similarly, for events with six or more clusters, 
we consider each pair of non-overlapping \piz candidates,
and perform a set of five-constraint fits under the $\KS\KL\piz\piz\gamma$ 
hypothesis. 
Both $\piz$ masses are constrained, and we retain the $\piz\piz\KL$
candidate combination yielding the lowest value of \chiKKppg.

Finally, we perform similar fits for all the other simulated signal
and background processes discussed in Sec.~\ref{sec:babar}, 
giving us additional \chisq variables that can be used to
select (or suppress) these processes.
\begin{figure}[tbh]
\begin{center}
\includegraphics[width=0.9\linewidth]{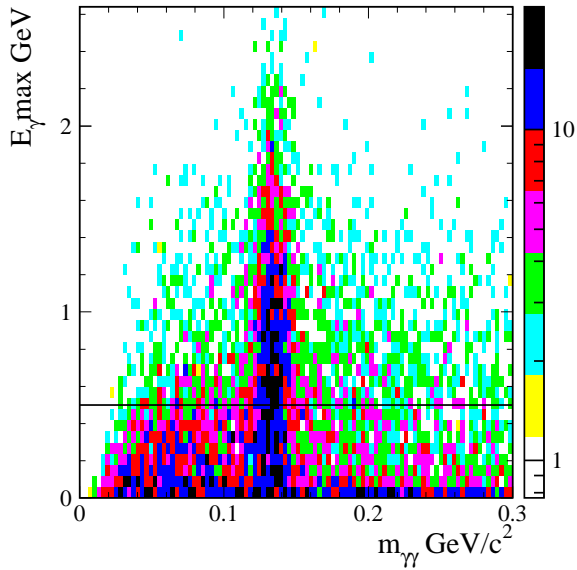}
\vspace{-0.4cm}
\caption{
Two-dimensional distribution 
of the higher cluster energy in a photon-candidate pair
vs.\ the corresponding diphoton mass $m_{\gamma\gamma}$ 
for all pairs of EMC clusters in $\KKpz\gamma$ events containing
neither the ISR photon, the $\KL$ candidate,
nor either photon in the $\piz$ candidate.
 }
\label{egammax}
\end{center}
\end{figure}

\section{\boldmath The $\KKpz$ final state}
\label{sec:ksklpiz}
\begin{figure*}[tb]
\begin{center}
\includegraphics[width=0.33\linewidth]{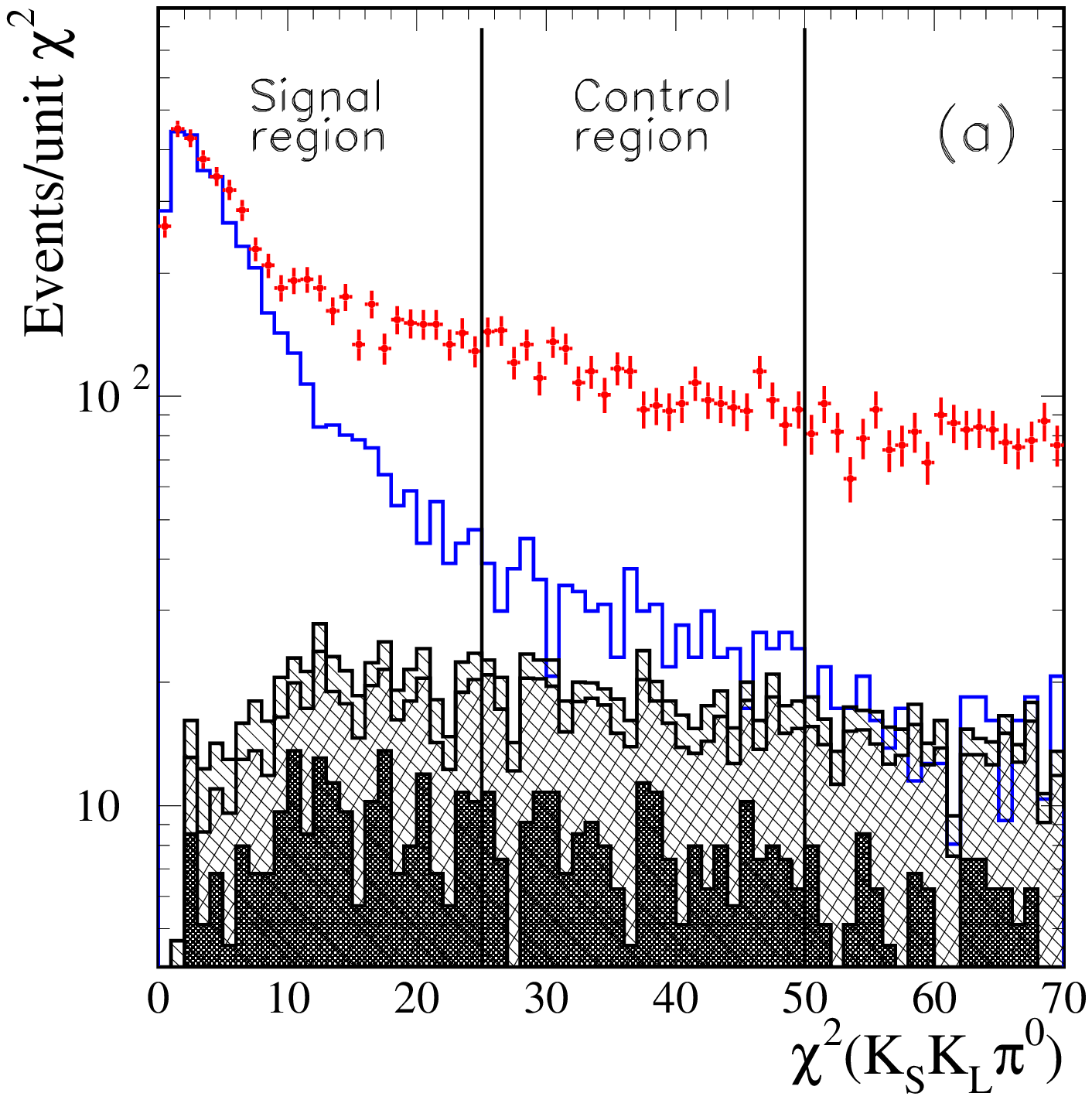}
\includegraphics[width=0.33\linewidth]{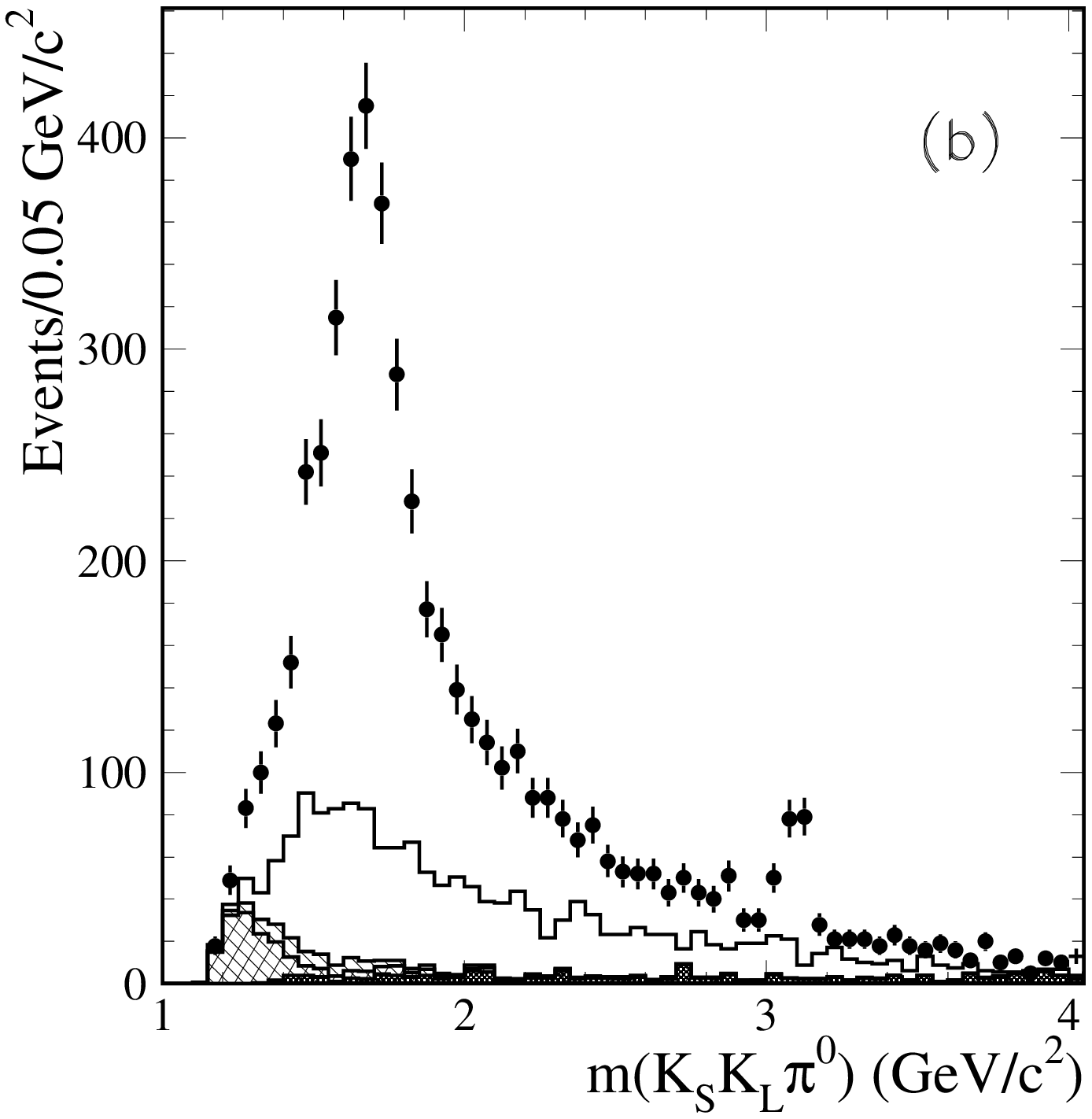}
\includegraphics[width=0.33\linewidth]{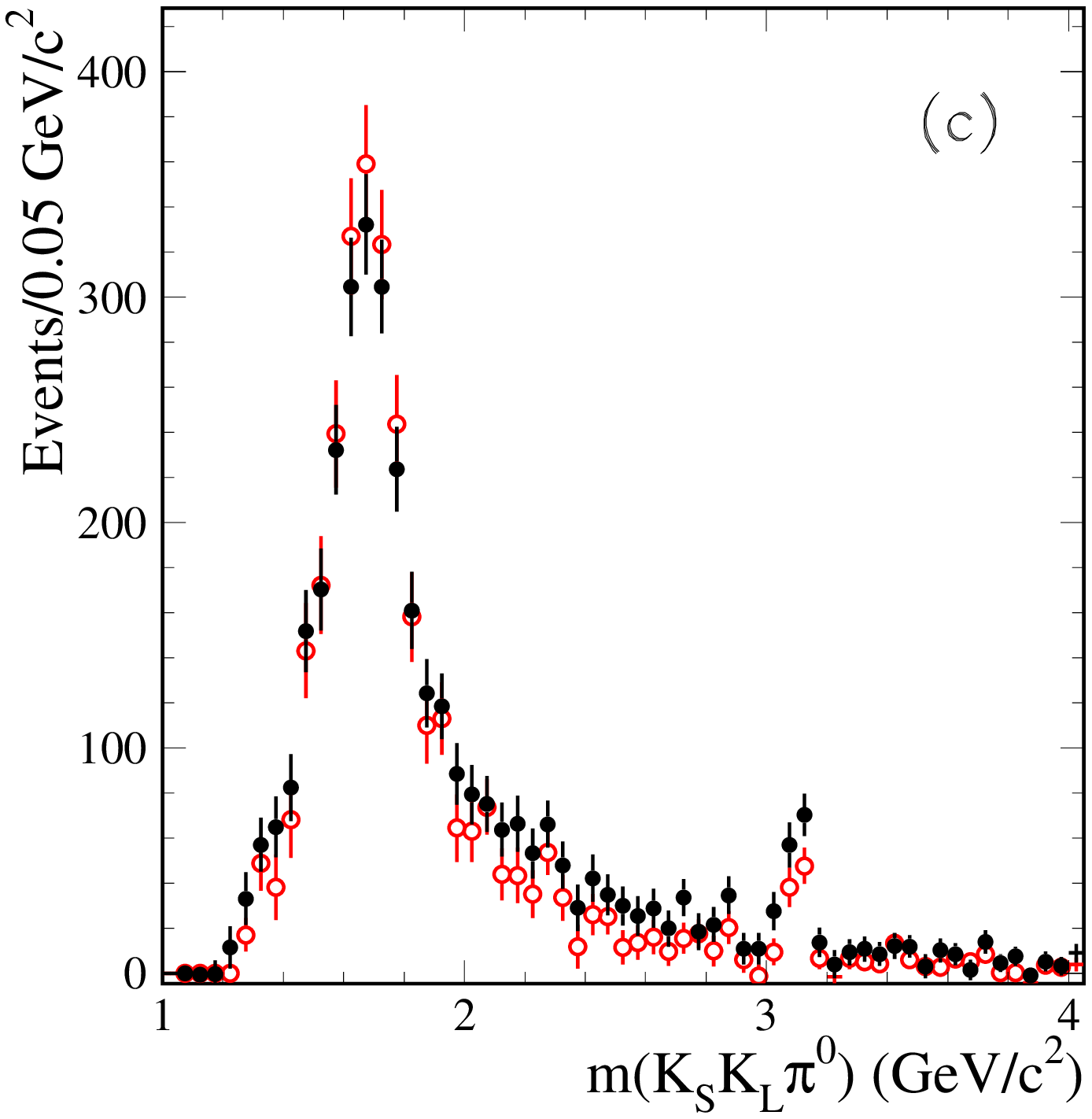}
\vspace{-0.5cm}
\caption{
  (a) The four-constraint \chisq distributions for data (points) and
      MC-simulated $\KKpz\gamma$ events (open histogram).
      The shaded, cross-hatched, and hatched areas represent the
      simulated backgrounds from  
      non-ISR $q\bar q$, ISR $\phi$, and ISR $\phi\eta$ events, respectively. 
  (b) The \KKpz invariant mass distribution for data events in the
      signal region (points).
      The shaded, cross-hatched, and hatched areas represent the simulated
      contributions from 
      non-ISR $q\bar q$, ISR $\phi$, and ISR $\phi\eta$ events, respectively,
      while the open histogram represents the total background
      estimated from the control region.  
  (c) The \KKpz invariant mass distribution after background
      subtraction (points).
      The open circles represent the contribution from the resonant
      process $\epem\to K^*(892)^0 \Kbar^0+\rm{c.c.}\to\KKpz$ (see text). 
}
\label{ksklpi0_chi2}
\end{center}
\end{figure*}

\subsection{Additional selection criteria}
\label{addksklpiz}
For the $\KS\KL\piz$ final state,
a few additional selection criteria are applied.
Considering all pairs of EMC clusters not assigned to the ISR photon,
$\piz$, or \KL candidates,
we observe a large signal from extra $\piz$'s, shown in Fig.~\ref{egammax}. 
It is especially strong when one of the clusters has high energy,
so we require $\rm E_{\gamma}^{max} < 0.5~\gev$.
This reduces backgrounds from several sources
with a loss of 3\% in simulated signal efficiency.
However, many ISR $\phi\gamma$ events with a false $\piz$, 
formed by accidential photons, remain.
To reduce this background, 
we require $\chiKKg>15$ if the fitted $\KS\KL$ invariant mass
$m(\KS\KL)$ is smaller than 1.04~\gevcc.

The 4C \chisq distribution for the remaining events under the 
$\KKpz\gamma$ hypothesis is shown as the points in  
Fig.~\ref{ksklpi0_chi2}(a), 
with the corresponding distribution for MC-simulated pure
$\KKpz\gamma$ events shown as the open histogram.
Both distributions are broader than typical 4C \chisq distributions
due to higher-order ISR,
which is present in both data and simulation, but not taken into
account in the fit.
The reliability of the simulated distribution has been demonstrated in
our previous studies, and is discussed below.
In the figure,
the simulated signal distribution is normalized to the data in the region
$\chiKKpg < 3$, where the contribution of higher-order
ISR is small and the background contamination is lowest, 
but still amounts to about 5\% of the signal.
The difference between the two distributions at high values gives an
indication of the level of background.

We define a signal region $\chiKKpg<25$ 
and a control region $25<\chiKKpg<50$ 
(vertical lines in Fig.~\ref{ksklpi0_chi2}(a)),
from which we estimate backgrounds in the signal region. 
The signal region contains 5441 data and 3402 signal-MC events, 
while the control region contains 2733 and 632 events, respectively.

\subsection{Background subtraction}
\label{sec:ksklpi0bkg}
We estimate known ISR backgrounds from simulation, and normalize the
simulated non-ISR background using the \piz peak, as described in  
Ref.~\cite{isrkskl}. 
The largest backgrounds we can evaluate in this way are shown in 
Fig.~\ref{ksklpi0_chi2}(a):
the shaded, cross-hatched, and hatched areas represent the simulated
backgrounds from non-ISR $q\bar q$, ISR $\KS \KL (\phi)$, and ISR
$\phi\eta$ events, respectively.  
The shapes of these three distributions are consistent with each other
and quite distinct from that expected for signal events.
However, these backgrounds account for less than half of the
observed difference between data and simulation at large \chisq values.
We assume the remaining background is from other ISR processes, 
with a \chiKKpg distribution similar in shape to those shown.

To obtain any distribution of the \KKpz signal events, 
we use the control region to estimate the sum of all backgrounds, 
following the procedure described in detail in Ref.~\cite{isrkskl}. 
In each bin of the distribution in question,
the background contribution is estimated as the difference between the
numbers of data and signal-MC events in the control region (see
Fig.~\ref{ksklpi0_chi2}(a)), 
normalized to the corresponding difference in the signal region.

The \KKpz invariant mass distribution of events in the
signal region 
is shown in Fig.~\ref{ksklpi0_chi2}(b) as the points.
The shaded, cross-hatched, and hatched areas represent the same
simulated backgrounds as in Fig.~\ref{ksklpi0_chi2}(a). 
The sum of all backgrounds, estimated from the control region,
is shown as the open histogram in Fig.~\ref{ksklpi0_chi2}(b),
and the extracted mass distribution for $\epem \to \KKpz$ signal events is
shown as the filled points in Fig.~\ref{ksklpi0_chi2}(c).   
We observe 3669 events in the mass range from threshold to 4.0~\gevcc.
In addition to a main peak around 1.8~\gevcc, a \jpsi signal is visible.

This procedure relies on good agreement between data and simulation in
the \chiKKpg distribution.
Considering our previous studies of \chisq distributions~\cite{isrkskl},
along with simulation and normalization statistics,
we estimate the relative systematic uncertainty 
on the background to be 30\%.
This results in an uncertainty on the background-subtracted signal of
about 10\% for $m(\KKpz)<2.2$~\gevcc,
increasing roughly linearly with mass to about 40\% at 3.2~\gevcc and
above. 

\begin{figure}[tbh]
\begin{center}
\includegraphics[width=0.9\linewidth]{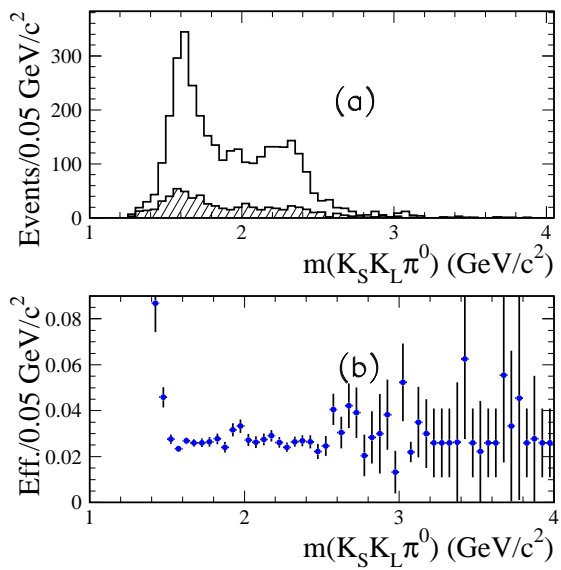}
\vspace{-0.4cm}
\caption{
(a) The reconstructed \KKpz invariant mass distribution for
    MC-simulated signal events in the signal (open histogram) 
    and control (hatched) regions of Fig.~\ref{ksklpi0_chi2}(a).
(b) The net reconstruction efficiency from the simulation.
}
\label{ksklpi0_eff}
\end{center}
\end{figure}

\subsection{Detection efficiency }
\label{sec:ksklpi0eff}
The selection procedures applied to the data are also applied to the         
MC-simulated event sample. 
The resulting distribution of the reconstructed \KKpz invariant mass
is shown in Fig.~\ref{ksklpi0_eff}(a) for events with \chiKKpg in the
signal (open histogram) and control (hatched) regions.
The reconstruction efficiency as a function of mass is obtained by
dividing the number of reconstructed MC events in each 50~\mevcc 
mass interval by the number generated in that interval,
and is shown in Fig.~\ref{ksklpi0_eff}(b). 
The effects of detector resolution, about 25~\mevcc, are included in
this efficiency.
Below 1.5~\gevcc the efficiency becomes very large, 
due to the rapidly changing cross section near threshold.  
Since the resolution is measured from the data and the shape of the
threshold rise is well simulated, we apply no correction. 
Nevertheless,
the backgrounds and the resolution-effect uncertainties are high in this region,
so we do not quote a cross section measurement below 1.4~\gevcc,
and we assign an additional 50\% (30\%) relative systematic
uncertainty for the mass bin at 1.425(1.475)~\gevcc.

This efficiency is corrected for the data-MC differences evaluated in
our previous studies.
The ISR photon detection efficiency has been studied using $\mu^+\mu^-\gamma$
events~\cite{isr2pi}, 
and we apply a polar-angle-dependent correction of typically
$-$1.5$\pm$0.5\% to the simulated efficiency.
The \KS detection efficiency has been studied very carefully at \babar,
with data-MC differences in the efficiency determined as a function of
the \KS direction and momentum. 
We apply a correction event by event, 
which introduces an overall correction of $+$1.1$\pm$1.0\% to the
efficiency.
The $\piz$ reconstruction efficiency has been studied in \babar\ using 
$\omega\gamma$ and $\omega\piz\gamma$ events, 
and the correction is found to be ($-$3$\pm$1)\%. 
The \KL detection requires a ($-$6.1$\pm$0.6)\% correction~\cite{isrkskl}. 
In total, there is a ($-$9.5$\pm$1.6)\% correction; 
this systematic uncertainty is small compared with that due to the
backgrounds, described above.

\begin{table*}
\caption{Summary of the $\epem\to K_S K_L\piz$ 
cross section measurement. Uncertainties are statistical only.}
\label{ksklpi0_tab}
\begin{ruledtabular}
\begin{tabular}{ c c c c c c c c }
$E_{\rm c.m.}$ (GeV) & $\sigma$ (nb)  
& $E_{\rm c.m.}$ (GeV) & $\sigma$ (nb) 
& $E_{\rm c.m.}$ (GeV) & $\sigma$ (nb) 
& $E_{\rm c.m.}$ (GeV) & $\sigma$ (nb)  
\\
\hline

  1.425 &  0.28 $\pm$  0.07 &  2.075 &  0.55 $\pm$  0.11 &  2.725 &  0.12 $\pm$  0.05 &  3.375 &  0.03 $\pm$  0.04 \\
  1.475 &  0.94 $\pm$  0.15 &  2.125 &  0.44 $\pm$  0.09 &  2.775 &  0.12 $\pm$  0.08 &  3.425 &  0.02 $\pm$  0.01 \\
  1.525 &  1.68 $\pm$  0.22 &  2.175 &  0.42 $\pm$  0.09 &  2.825 &  0.10 $\pm$  0.06 &  3.475 &  0.05 $\pm$  0.03 \\
  1.575 &  2.60 $\pm$  0.28 &  2.225 &  0.36 $\pm$  0.08 &  2.875 &  0.15 $\pm$  0.10 &  3.525 &  0.01 $\pm$  0.03 \\
  1.625 &  2.89 $\pm$  0.26 &  2.275 &  0.48 $\pm$  0.09 &  2.925 &  0.04 $\pm$  0.03 &  3.575 &  0.04 $\pm$  0.03 \\
  1.675 &  3.15 $\pm$  0.30 &  2.325 &  0.31 $\pm$  0.07 &  2.975 &  0.10 $\pm$  0.10 &  3.625 &  0.03 $\pm$  0.03 \\
  1.725 &  2.79 $\pm$  0.29 &  2.375 &  0.18 $\pm$  0.07 &  3.025 &  0.06 $\pm$  0.03 &  3.675 &  0.00 $\pm$  0.01 \\
  1.775 &  1.96 $\pm$  0.23 &  2.425 &  0.26 $\pm$  0.07 &  3.075 &  0.31 $\pm$  0.09 &  3.725 &  0.04 $\pm$  0.04 \\
  1.825 &  1.30 $\pm$  0.18 &  2.475 &  0.25 $\pm$  0.08 &  3.125 &  0.24 $\pm$  0.12 &  3.775 &  0.01 $\pm$  0.01 \\
  1.875 &  1.12 $\pm$  0.18 &  2.525 &  0.19 $\pm$  0.06 &  3.175 &  0.05 $\pm$  0.04 &  3.825 &  0.03 $\pm$  0.02 \\
  1.925 &  0.79 $\pm$  0.12 &  2.575 &  0.09 $\pm$  0.04 &  3.225 &  0.02 $\pm$  0.03 &  3.875 &  0.00 $\pm$  0.01 \\
  1.975 &  0.55 $\pm$  0.10 &  2.625 &  0.14 $\pm$  0.05 &  3.275 &  0.04 $\pm$  0.03 &  3.925 &  0.02 $\pm$  0.02 \\
  2.025 &  0.58 $\pm$  0.11 &  2.675 &  0.07 $\pm$  0.03 &  3.325 &  0.05 $\pm$  0.04 &  3.975 &  0.01 $\pm$  0.01 \\

\end{tabular}
\end{ruledtabular}
\end{table*}

\begin{figure}[tbh]
\begin{center}
\includegraphics[width=0.9\linewidth]{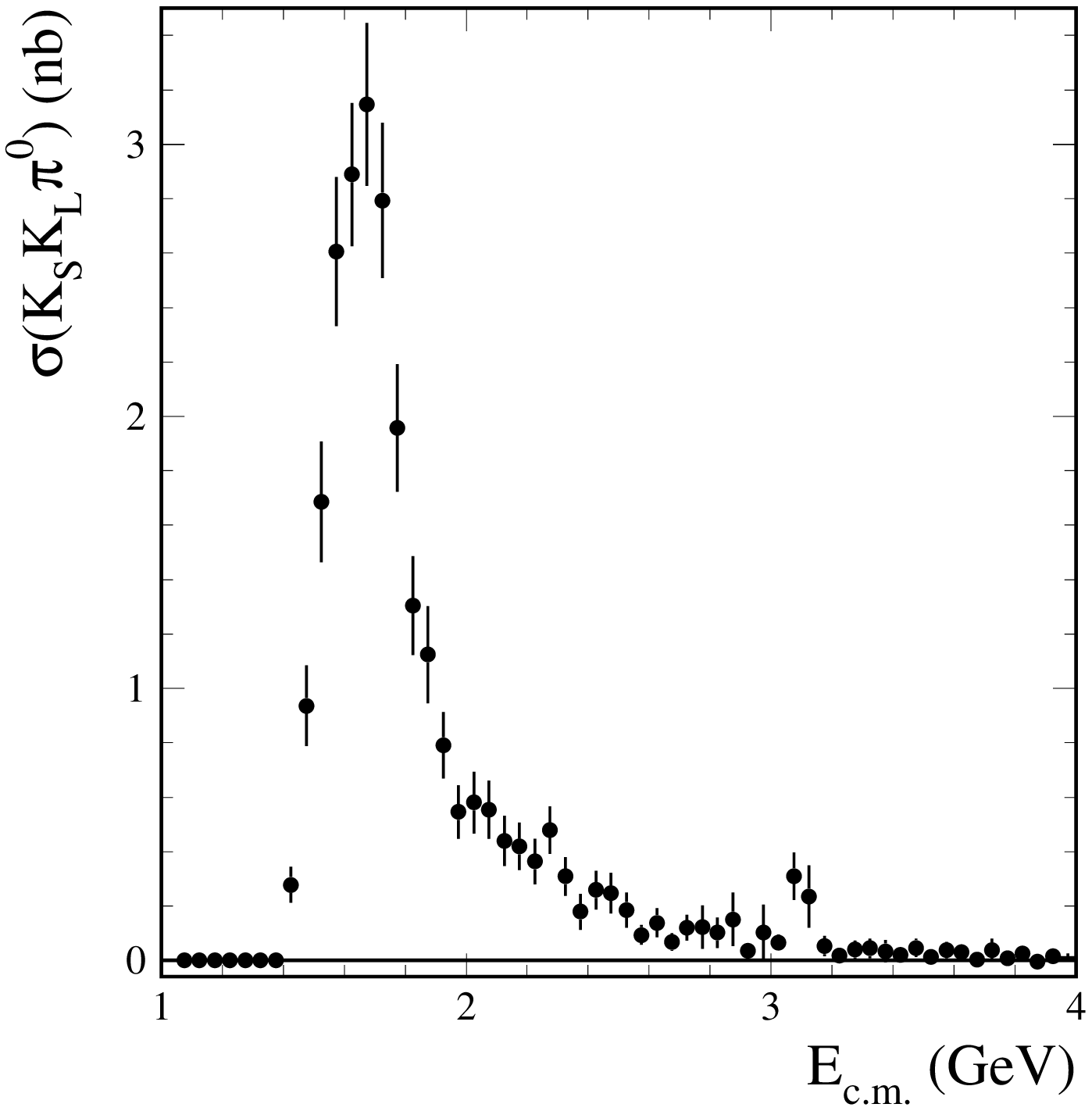}
\vspace{-0.4cm}
\caption{
The $\epem\to \KKpz$ cross section.
The error bars are statistical only.
}
\label{ksklpi0_xs}
\end{center}
\end{figure}
\subsection{\boldmath The $\epem\to \KKpz$ cross section }
\label{sec:xsksklpi0}
The cross section for \epem annihilation into \KKpz is
calculated from 
\begin{equation}
  \sigma(\KKpz)(\Ecm)
  = \frac{dN_{\KKpz\gamma}(\Ecm)}
         {d{\cal L}(\Ecm)
          \cdot\epsilon(\Ecm)
          \cdot R }  ,
\label{xsformular}
\end{equation}
where $\Ecm\equiv m(\KKpz)$; 
$dN_{\KKpz\gamma}$ is the number of selected, background-subtracted
\KKpz events in the interval $dE_{\rm c.m.}$; 
$\epsilon(\Ecm)=\epsilon^{\rm MC}(\Ecm)\cdot(1+\delta_{\rm corr})$
is the simulated detection efficiency corrected for data-MC
differences, as described above.
The radiative correction $R$ is unity within 1\%, 
with an estimated precision of about 1\%.
The differential ISR luminosity $d{\cal L}(\Ecm)$ associated with
the interval $d\Ecm$ centered at an effective collision
energy of $\Ecm$ is calculated using the leading order formula
(see, for example, Ref.~\cite{isr3pi}),
and the systematic uncertainty associated with the luminosity
determination is estimated to be 0.5\%.

The cross section is shown as a function of energy in
Fig.~\ref{ksklpi0_xs}, 
and listed in Table~\ref{ksklpi0_tab}. 
There are no previous measurements for this final state. 
We do not quote the cross section 
from threshold (1.13~\gev) to 1.4~\gev, where it
shows a sharp rise to a maximum value of about 3~nb near 1.7~\gev, 
presumably dominated by the $\phi(1680)$ resonance, 
and a slow decrease toward higher energies, 
perturbed by the \jpsi signal.
Only statistical uncertainties are shown.
The systematic uncertainty is dominated by background contributions,
and amounts to about 10\%  near the peak of the cross section
(1.7~\gev), 
increasing roughly lineary with decreasing cross section to about 30\%
in the 2.5--3~\gev region, 
always similar in size to the statistical uncertainty.
Above the \jpsi mass, statistics dominate the $\sim$40\% systematic
uncertainty. 

\begin{figure}[tbh]
\begin{center}
\includegraphics[width=0.98\linewidth]{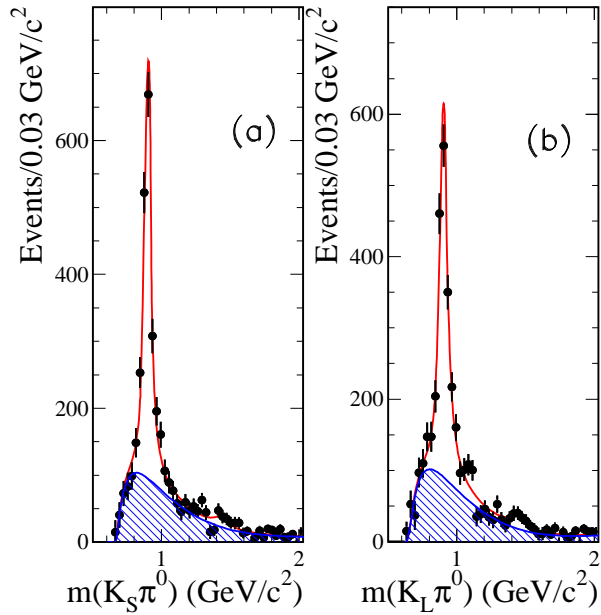}
\vspace{-0.4cm}
\caption{
The background-subtracted (a) $\KS\piz$ and (b) $\KL\piz$ invariant
mass distributions in $\epem\to\KKpz$ events. 
The curves represent the results of the fits described in the text,
with the hatched areas representing the non-resonant components.
}
\label{kstarpi}
\end{center}
\end{figure}
\subsection{\boldmath The $K^{*}(892)^0$ and $K_2^{*}(1430)^0$
  contributions } 
\label{sec:kstar}
Figure~\ref{kstarpi} shows the distributions of the fitted $\KS\piz$
and $\KL\piz$ invariant masses in the selected \KKpz events, 
after background subtraction.  
Clear signals corresponding to the $K^{*}(892)^0$ resonance are visible, 
as well as indications of $K_2^{*}(1430)^0$ production.

We fit these distributions with a sum of two incoherent Breit-Wigner functions
and a function describing the non-resonant contribution,
yielding $1750\pm84$ $K^{*}(892)^0\to \KS\piz$ decays,
$1795\pm56$ $K^{*}(892)^0\to \KL\piz$ decays,
and a total of $145\pm54$ $K_2^{*}(1430)^0$ decays.
The sum of these $K^{*0}$ decays is consistent with the
total number of \KKpz events, 
indicating that the process is dominated by 
$K^{*0} \Kbar^0 + c.c.$, 
and mostly $K^{*}(892)^0 \Kbar^0 + c.c.$, production. 

Indeed, if we perform fits similar to those shown in Fig.~\ref{kstarpi} for
events in each 0.05~\gevcc interval of the \KKpz invariant mass, 
and sum the $K^*(892)^0\KS$ and $K^*(892)^0\KL$ yields,
we obtain the \KKpz mass distribution shown in
Fig.~\ref{ksklpi0_chi2}(c) by the open circles.
The errors are statistical only, and
the difference between the number of \KKpz events and the
$K^*(892)^0\Kbar^0$ contribution in each bin is less than the 
systematic uncertainty due to the background subtraction procedure.

\begin{figure}[tbh]
\begin{center}
\includegraphics[width=\linewidth]{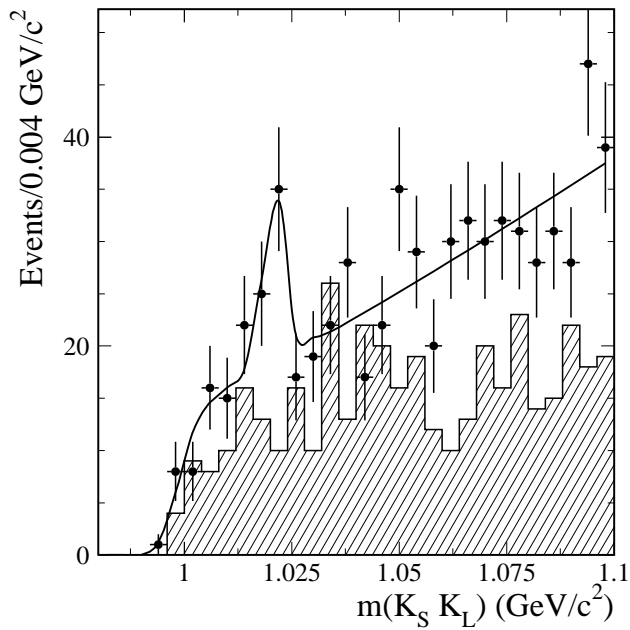}
\vspace{-0.4cm}
\caption{
  The $\KS\KL$ invariant mass distribution in selected \KKpz events
  (dots), 
  and the background estimated from the control region (hatched
  histogram) of Fig.~\ref{ksklpi0_chi2}(a).
  The solid line represents the result of the fit
  described in the text. 
}
\label{phipi0}
\end{center}
\end{figure}
\begin{figure}[tbh]
\begin{center}
\includegraphics[width=\linewidth]{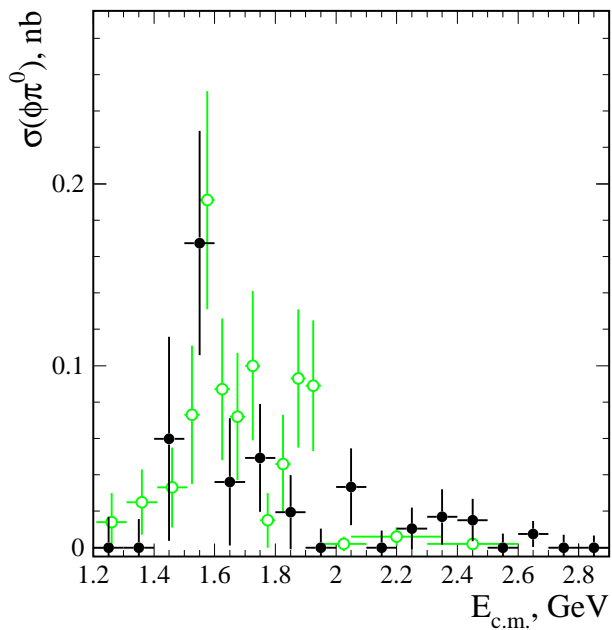}
\vspace{-0.4cm}
\caption{
The $\epem\to\phi\piz$ cross section from this work (dots) compared
with that obtained in the $\Kp\Km\piz$ channel~\cite{isrkkpi} (circles).
The error bars are statistical only.
}
\label{xs_phipi0}
\end{center}
\end{figure}
\subsection{\boldmath The $\phi(1020)\piz$ contribution } 
\label{sec:phipi0}
Figure~\ref{phipi0} shows the distribution of the fitted $\KS \KL$
invariant mass for the selected \KKpz events before background
subtraction (dots), 
along with the background (histogram) estimated from the \chiKKpg
control region.
A $\phi(1020)$ signal is visible in the $\KS\KL\piz$ signal, but not
in the background.
A fit with a Gaussian plus polynomial function yields a total of
$29\pm 9$ $\phi \to \KS\KL$ decays in $\KS\KL\piz$ events.

Fitting the $m(\KS \KL)$ distribution in 0.1~\gevcc bins of the $\KS\KL\piz$
mass, 
we obtain a $\phi\piz$ invariant mass spectrum for the $\KS\KL\piz$
final state.
Using Eq.~\ref{xsformular} and  
the $\phi\to\KS\KL$ branching fraction~\cite{PDG},
we calculate a cross section for this intermediate state, shown in
Fig.~\ref{xs_phipi0} (dots) and listed in Table~\ref{phipi0_tab}. 
Only statistical uncertainties are shown.  
The systematic uncertainties of 10--30\% relative in this mass range
are smaller than the statistical uncertainties.

The results are consistent with those observed in our previous study
of the $\Kp\Km\piz$ final state~\cite{isrkkpi}, 
also shown in Fig.~\ref{xs_phipi0} (circles).
Together, our measurements suggest a possible resonant structure near
1.6~\gevcc with isospin $I=1$. 
The low cross section is expected, as the $\phi\piz$ channels are
suppressed by the OZI-rule.

\begin{table}
\caption{Summary of the $\epem\to \phi\piz$ 
cross section measurement. Uncertainties are statistical only.}
\label{phipi0_tab}
\begin{ruledtabular}
\begin{tabular}{ c c c c }
$E_{\rm c.m.}$ (GeV) & $\sigma$ (nb)  
& $E_{\rm c.m.}$ (GeV) & $\sigma$ (nb) 

\\
\hline

 1.25 & 0.00 $\pm$ 0.02 & 2.15 & 0.00 $\pm$ 0.01\\
 1.35 & 0.00 $\pm$ 0.02 & 2.25 & 0.01 $\pm$ 0.01\\
 1.45 & 0.06 $\pm$ 0.06 & 2.35 & 0.02 $\pm$ 0.02\\
 1.55 & 0.17 $\pm$ 0.06 & 2.45 & 0.02 $\pm$ 0.01\\
 1.65 & 0.04 $\pm$ 0.03 & 2.55 & 0.00 $\pm$ 0.01\\
 1.75 & 0.05 $\pm$ 0.03 & 2.65 & 0.01 $\pm$ 0.01\\
 1.85 & 0.02 $\pm$ 0.02 & 2.75 & 0.00 $\pm$ 0.01\\
 1.95 & 0.00 $\pm$ 0.01 & 2.85 & 0.00 $\pm$ 0.01\\
 2.05 & 0.03 $\pm$ 0.02 &      &                \\

\end{tabular}
\end{ruledtabular}
\end{table}

\section{\boldmath The $\KKeta$ final state}
\label{sec:kskketa}
\begin{figure*}[tb]
\begin{center}
\includegraphics[width=0.33\linewidth]{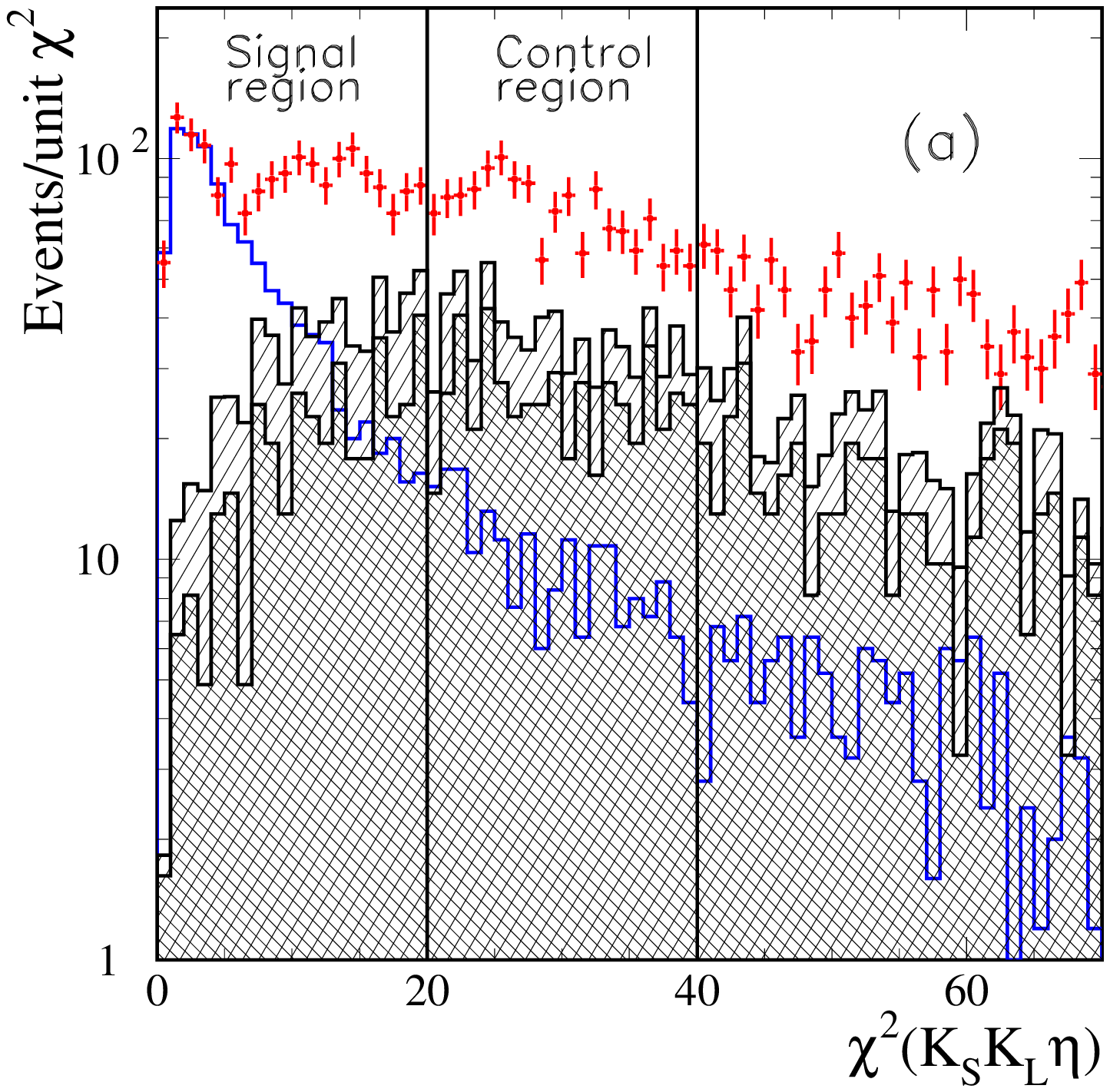}
\includegraphics[width=0.33\linewidth]{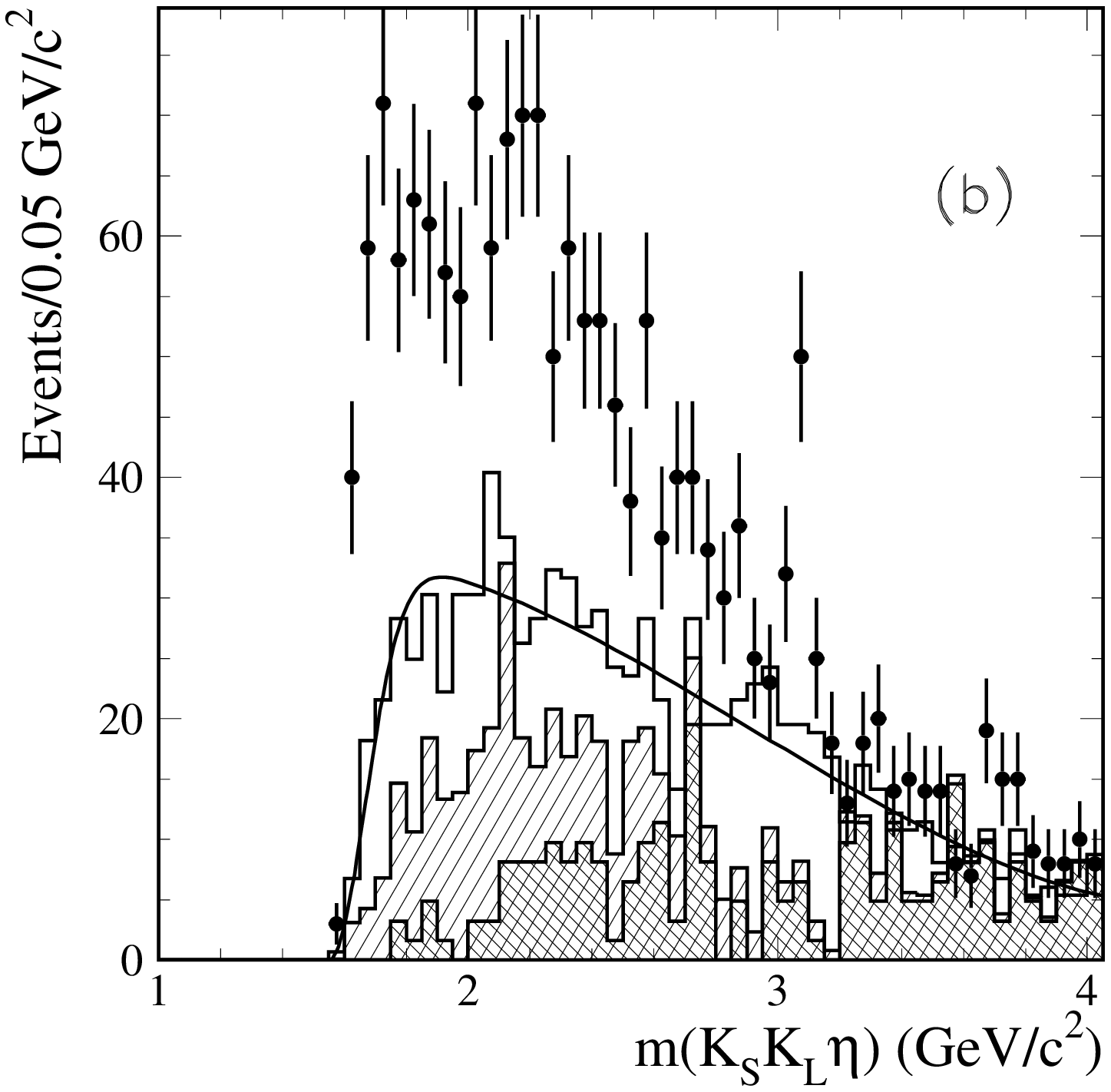}
\includegraphics[width=0.33\linewidth]{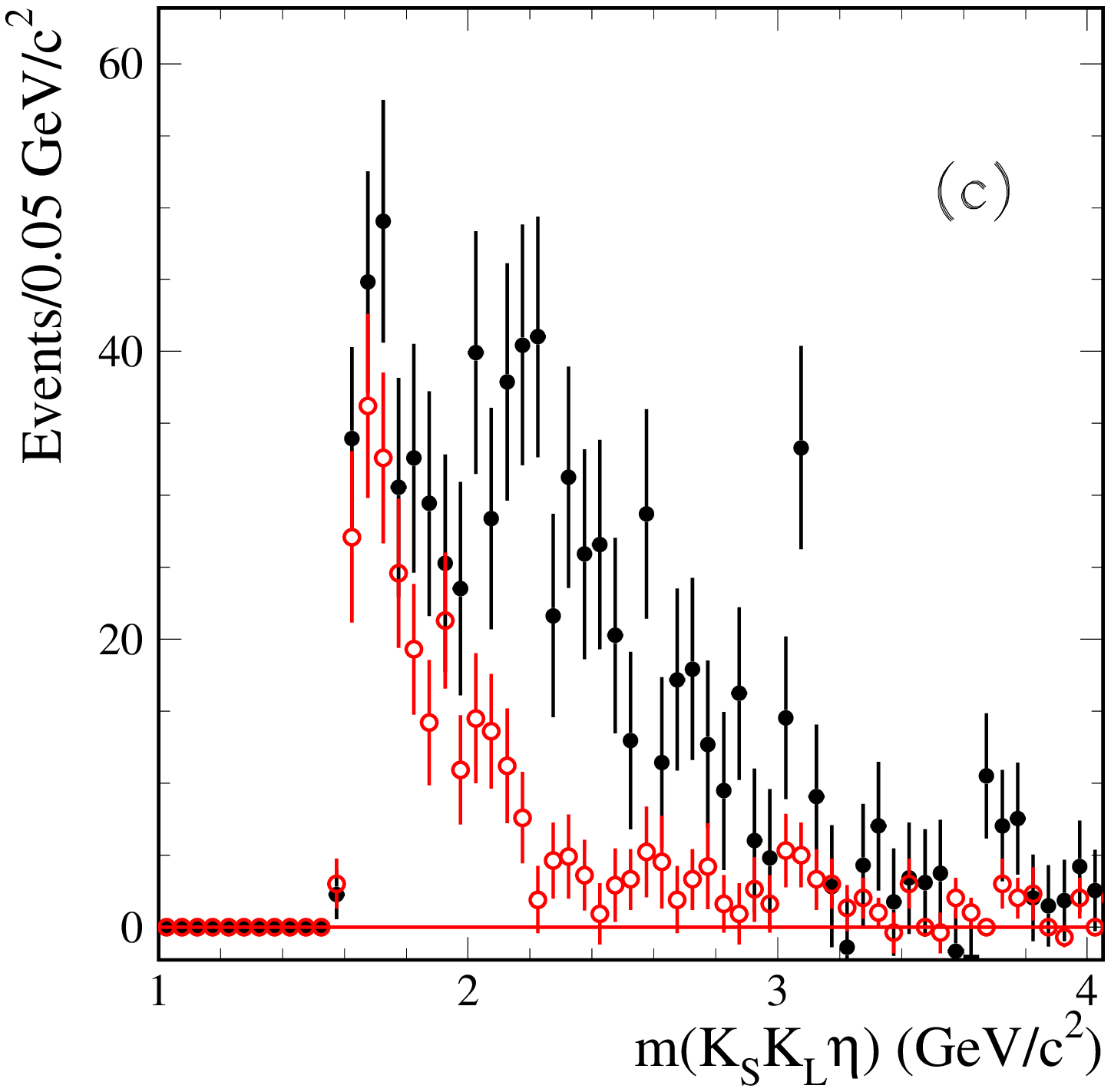}
\vspace{-0.5cm}
\caption{
  (a) The four-constraint \chisq distributions for data (points) and
      MC-simulated $\KKeta\gamma$ events (open histogram).
      The cross-hatched and hatched areas represent the
      simulated backgrounds from non-ISR $q\bar q$ events, 
      and the sum of ISR \KKpz, $\KS\KL$, and $\KKppz$ events,
      respectively.
  (b) The \KKeta invariant mass distribution for data events in the
      signal region (points).
      The cross-hatched and hatched areas represents 
      simulated backgrounds from non-ISR $q\bar q$ and the sum of
      known ISR events, respectively,
      while the open histogram represents the total background,
      estimated from the control region.  
      The curve shows the empirical fit used for background subtraction.
  (c) The \KKeta invariant mass distribution after background
      subtraction (points).
      The open circles represent the contribution from the resonant
      process $\epem\to\phi\eta\to\KKeta$ (see text).
}
\label{kskleta_chi2}
\end{center}
\end{figure*}
\subsection{Final selection and backgrounds}
\label{addkskleta}
We apply the same requirements on extra \piz's and \chiKKg as for the
\KKpz final state (see Sec.~\ref{sec:ksklpiz}),
and consider the $\eta$-\KL combination in each event with the best
\chisq under the $\KKeta\gamma$ hypothesis.
Figure~\ref{kskleta_chi2}(a) shows the \chiKKeg distribution of the
remaining events in the data (dots) compared with that of the signal
simulation (open histogram).  
The simulated distribution is normalized to the data in the region
$\chiKKeg<3$, where the contribution of higher-order
ISR is small and the background contamination is lowest, 
but still amounts to about 10\% of the signal. 
The cross-hatched and hatched areas represent the
simulated contributions from non-ISR $q\bar q$ events 
and the sum of ISR \KKpz, ISR $\KS \KL$ , and ISR $\KKppz$ events,
respectively; 
together, they account for about half of the excess of data over
signal events at high values of \chisq.

We define a signal region $\chiKKeg<20$ 
and a control region $20<\chiKKeg<40$ 
(vertical lines in Fig.~\ref{kskleta_chi2}(a)),
containing 1829 data and 2518 signal-MC events, 
and 1473 data and 495 signal-MC events, respectively.
The $m(\KKeta)$ distribution for the events in the signal region
is shown in Fig.~\ref{kskleta_chi2}(b) as points,
along with the sum of the simulated background processes as the
cross-hatched and hatched areas.  
Using events from the control region (see Sec.~\ref{sec:ksklpi0bkg})
we calculate the total background contribution,
assumed to be dominated by ISR channels,
and show it as the open histogram in Fig.~\ref{kskleta_chi2}(b).

We fit the total background with a smooth function to reduce
fluctuations, 
and use the result (curve in Fig.~\ref{kskleta_chi2}(b)) for the
background subtraction. 
This yields a total of $864\pm43$ signal events with masses between
threshold and 4.0~\gevcc, 
with the mass distribution shown in Fig.~\ref{kskleta_chi2}(c). 
Again, 
we estimate the relative systematic uncertainty on the background as
30\%,
corresponding to an uncertainty on the cross section of about 15\% for
$m(\KKeta)<2.2$~\gevcc, 
increasing roughly lineary to 30\% at 3.0~\gevcc,
and over 100\% above 3.2~\gevcc.

\begin{figure}[tbh]
\begin{center}
\includegraphics[width=0.9\linewidth]{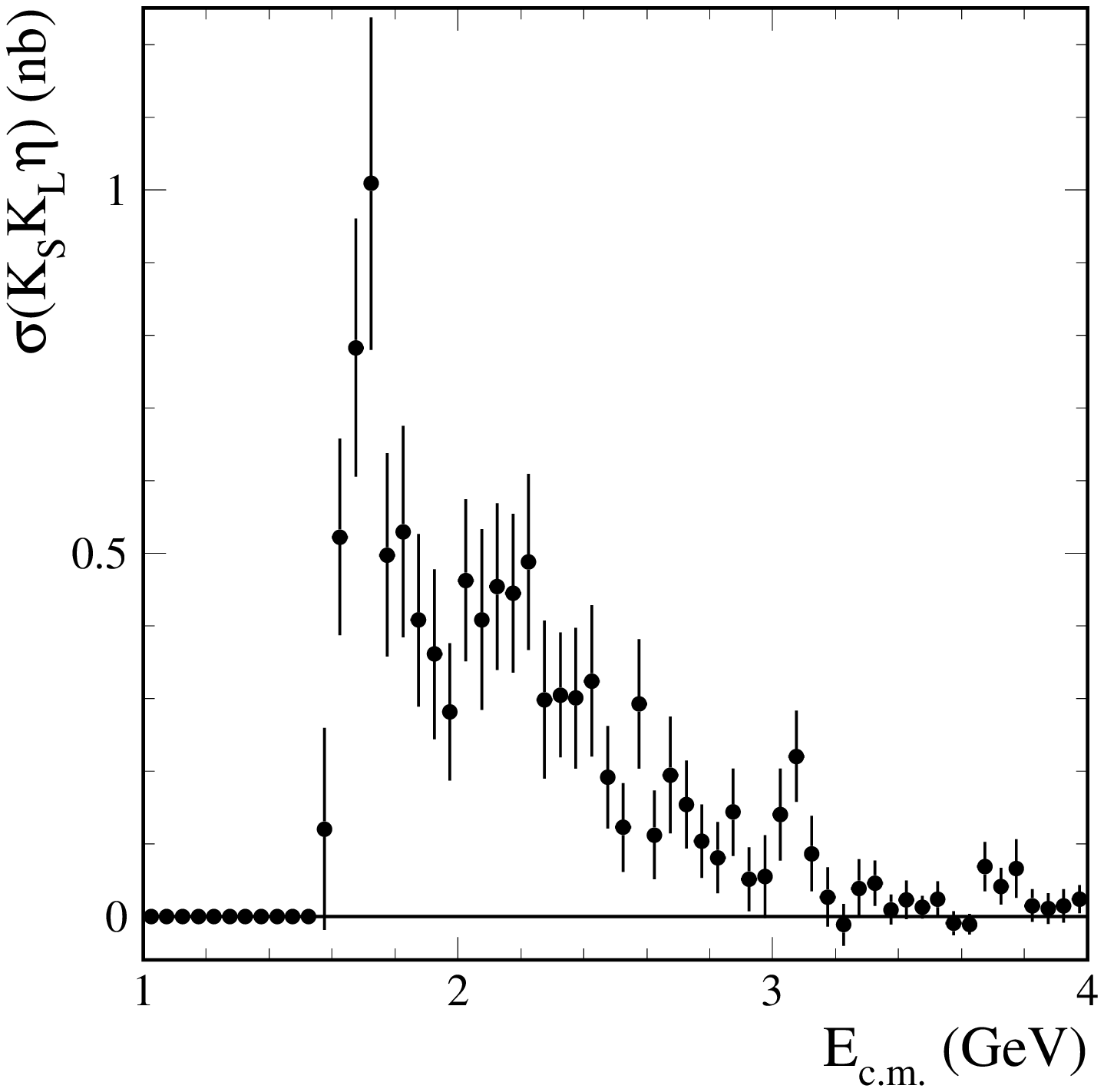}
\vspace{-0.4cm}
\caption{
The $\epem\to\KKeta$ cross section.
}
\label{kskleta_xs}
\end{center}
\end{figure}
\subsection{\boldmath Cross section for $\epem\to \KKeta$ }
\label{sec:xskskleta}
We calculate the $\epem \!\!\to\!\KKeta$ cross section as a
function of the effective c.m.\ energy using
Eq.~\ref{xsformular}.  
The simulated efficiency is 1.6\% and shows no dependence on the 
$\KKeta$ invariant mass. 
All efficiency corrections discussed in Sec.~\ref{sec:xsksklpi0} are
applied, 
in particular, the same correction is applied to the $\eta$
reconstruction efficiency as for the \piz.

The fully corrected cross section is shown in Fig.~\ref{kskleta_xs}
and listed in Table~\ref{kskleta_tab},
with statistical uncertainties only.
There are no other measurements for this final state. 
The cross section shows a steep rise from threshold at 1.6~\gev,
a maximum value of about 1 nb near 1.7~\gev,
and a decrease with increasing energy, 
punctuated by a clear \jpsi signal (discussed in Sec.~\ref{sec:jpsi}).
The relative systematic uncertainty is dominated by the uncertainty of the
backgrounds, 
totals 15\% at the peak of the cross section, 
increases roughly lineary to about 30\% at 3~\gev, 
and exceeds 100\% at higher energies.

\begin{figure}[tbh]
\begin{center}
\includegraphics[width=0.9\linewidth]{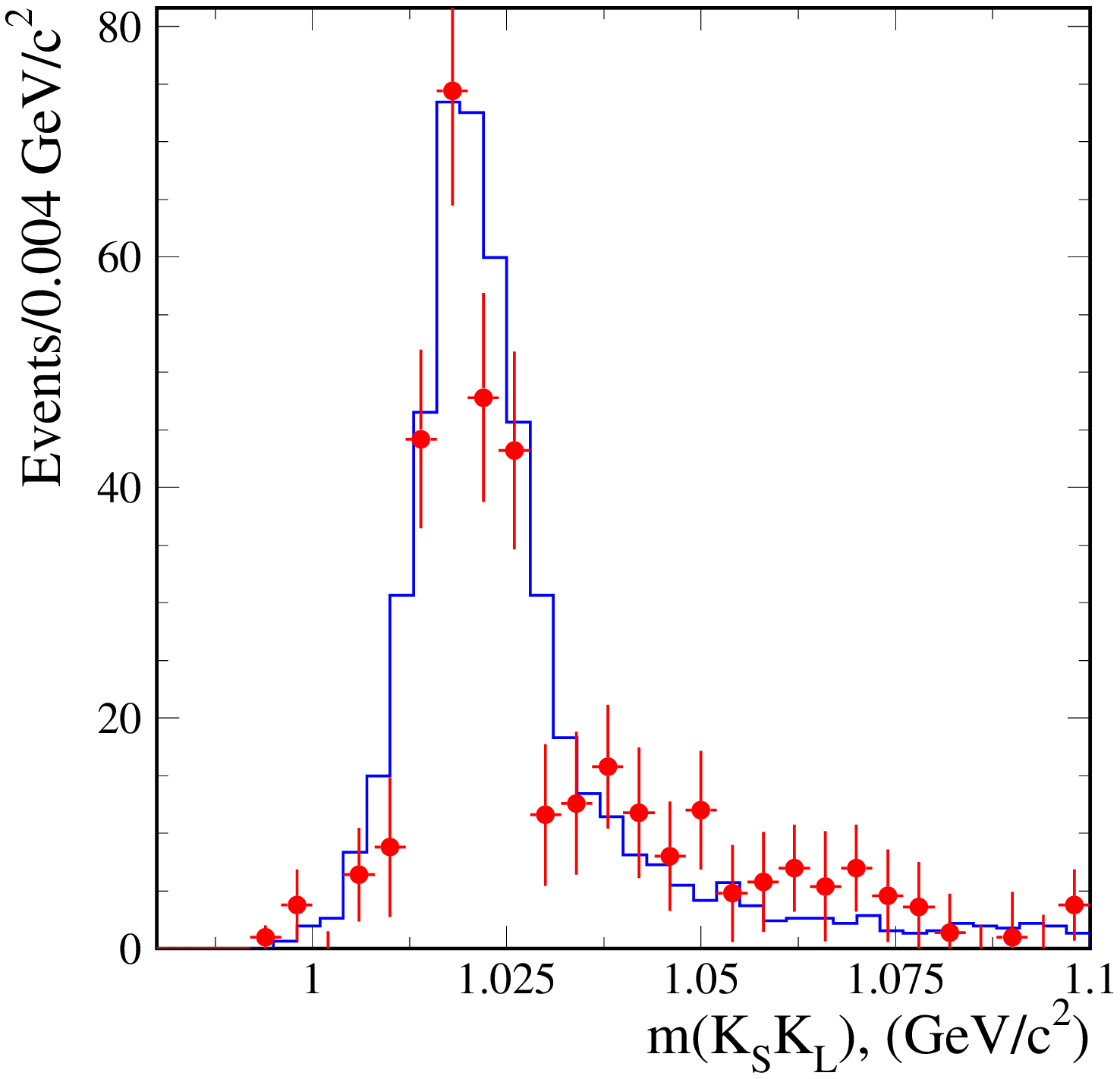}
\vspace{-0.4cm}
\caption{
The background-subtracted $\KS \KL$ invariant mass distribution in
$\epem\to\KKeta$ events (points),
compared with that of simulated $\epem\to\phi\eta$ events (histogram),
normalized to number of experimental events..
}
\label{phimass}
\end{center}
\end{figure}
\subsection{\boldmath The $\phi(1020)\eta$ contribution }
\label{sec:xsphieta}
Figure~\ref{phimass} shows the background-subtracted $\KS\KL$
invariant mass distribution in $\epem\to\KKeta$ events (dots),
compared with that of simulated ISR $\phi\eta$ events (histogram).  
The two distributions are consistent at low mass values,
and we simply take the number of events with $m(\KS\KL)<1.05$~\gevcc
as an estimate of the $\phi\eta$ contribution.
It totals $386\pm 20$ events, with the \KKeta invariant mass
distribution shown in Fig.~\ref{kskleta_chi2}(c) as the open circles.
The $\phi\eta$ channel dominates \KKeta production for masses below
about 2~\gevcc, 
but its contribution decreases rapidly for higher masses, 
and shows no significant \jpsi signal.

\begin{figure}[tbh]
\begin{center}
\includegraphics[width=0.9\linewidth]{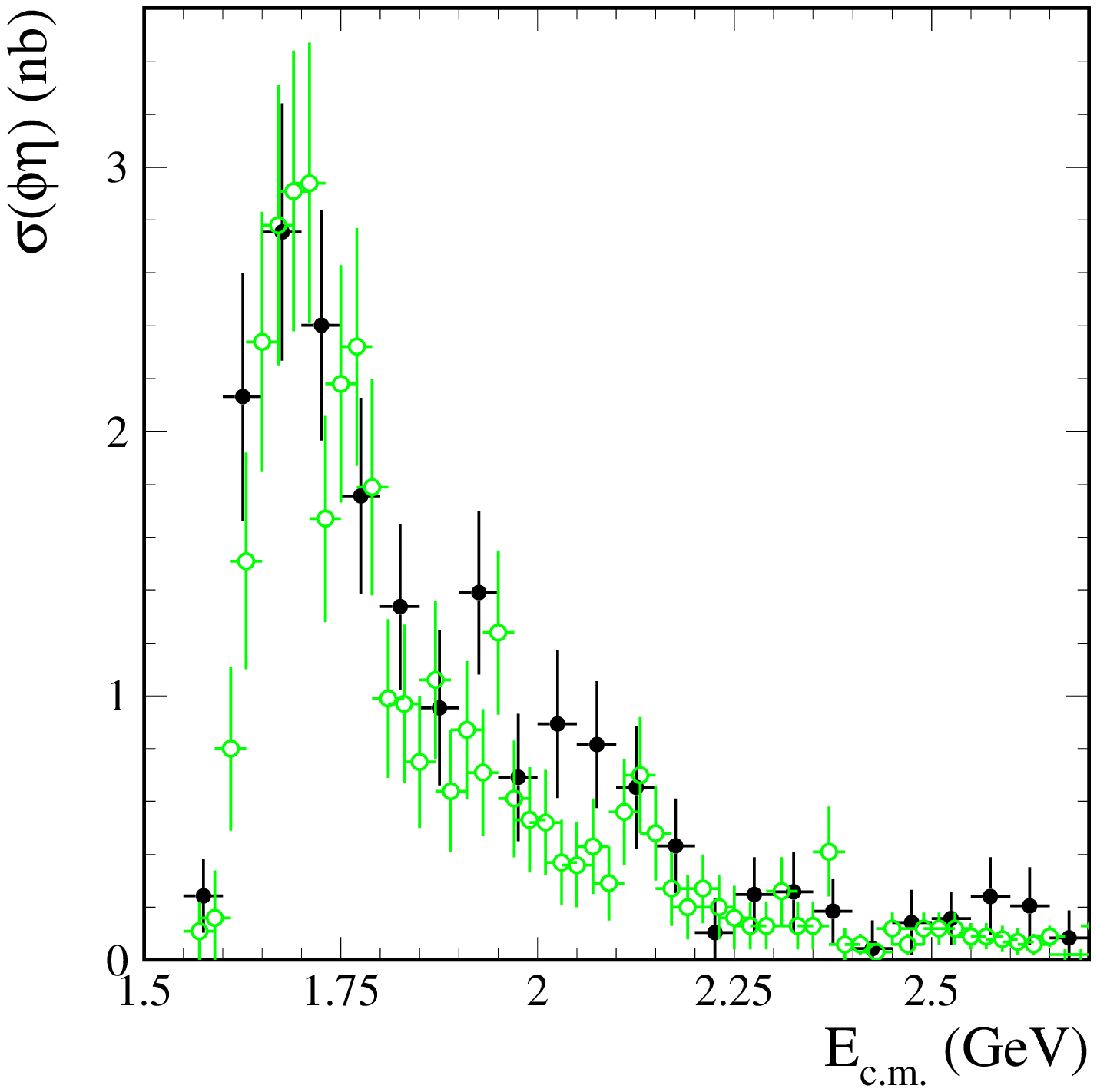}
\vspace{-0.4cm}
\caption{
The $\epem\to \phi(1020)\eta$ cross section obtained from this work
(solid dots),  
compared with the previous \babar\ measurement in the $\Kp\Km\eta$
final state~\cite{isrkkpi} (open circles). Only statistical errors are shown.
}
\label{phieta_xs}
\end{center}
\end{figure}

Using these events, we calculate the $\epem\to\phi\eta$ cross section,
which is shown in Fig.~\ref{phieta_xs} as the points.
It is consistent with our previous measurement~\cite{isrkkpi}
in the $\Kp\Km\eta$ final state (circles). 
Again, only statistical uncertainties are shown, 
and they are larger than the 15--30\% systematic uncertainties.
We observe no significant structures in the $\KS\eta$ or in the
$\KL\eta$ invariant mass distributions.

\begin{table*}
\caption{Summary of the $\epem\to K_S K_L\eta$ 
cross section measurement. Uncertainties are statistical only.}
\label{kskleta_tab}
\begin{ruledtabular}
\begin{tabular}{ c c c c c c c c }
$E_{\rm c.m.}$ (GeV) & $\sigma$ (nb)  
& $E_{\rm c.m.}$ (GeV) & $\sigma$ (nb) 
& $E_{\rm c.m.}$ (GeV) & $\sigma$ (nb) 
& $E_{\rm c.m.}$ (GeV) & $\sigma$ (nb)  
\\
\hline

        &                   &  2.075 &  0.41 $\pm$  0.12 &  2.725 &  0.15 $\pm$  0.06 &  3.375 &  0.01 $\pm$  0.02 \\
        &                   &  2.125 &  0.45 $\pm$  0.12 &  2.775 &  0.10 $\pm$  0.05 &  3.425 &  0.02 $\pm$  0.03 \\
        &                   &  2.175 &  0.44 $\pm$  0.11 &  2.825 &  0.08 $\pm$  0.05 &  3.475 &  0.01 $\pm$  0.02 \\
  1.575 &  0.12 $\pm$  0.14 &  2.225 &  0.49 $\pm$  0.12 &  2.875 &  0.14 $\pm$  0.06 &  3.525 &  0.02 $\pm$  0.02 \\
  1.625 &  0.52 $\pm$  0.14 &  2.275 &  0.30 $\pm$  0.11 &  2.925 &  0.05 $\pm$  0.04 &  3.575 & -0.01 $\pm$  0.02 \\
  1.675 &  0.78 $\pm$  0.18 &  2.325 &  0.31 $\pm$  0.09 &  2.975 &  0.05 $\pm$  0.06 &  3.625 & -0.01 $\pm$  0.01 \\
  1.725 &  1.01 $\pm$  0.23 &  2.375 &  0.30 $\pm$  0.10 &  3.025 &  0.14 $\pm$  0.06 &  3.675 &  0.07 $\pm$  0.03 \\
  1.775 &  0.50 $\pm$  0.14 &  2.425 &  0.32 $\pm$  0.10 &  3.075 &  0.22 $\pm$  0.06 &  3.725 &  0.04 $\pm$  0.03 \\
  1.825 &  0.53 $\pm$  0.14 &  2.475 &  0.19 $\pm$  0.07 &  3.125 &  0.09 $\pm$  0.05 &  3.775 &  0.07 $\pm$  0.04 \\
  1.875 &  0.41 $\pm$  0.12 &  2.525 &  0.12 $\pm$  0.06 &  3.175 &  0.03 $\pm$  0.04 &  3.825 &  0.01 $\pm$  0.02 \\
  1.925 &  0.36 $\pm$  0.12 &  2.575 &  0.29 $\pm$  0.09 &  3.225 & -0.01 $\pm$  0.03 &  3.875 &  0.01 $\pm$  0.02 \\
  1.975 &  0.28 $\pm$  0.09 &  2.625 &  0.11 $\pm$  0.06 &  3.275 &  0.04 $\pm$  0.04 &  3.925 &  0.01 $\pm$  0.02 \\
  2.025 &  0.46 $\pm$  0.11 &  2.675 &  0.19 $\pm$  0.08 &  3.325 &  0.05 $\pm$  0.03 &  3.975 &  0.02 $\pm$  0.02 \\

\end{tabular}
\end{ruledtabular}
\end{table*}

\section{\boldmath The \KKppz final state}
\label{sec:kskl2pi0}
\begin{figure*}[tb]
\begin{center}
\includegraphics[width=0.33\linewidth]{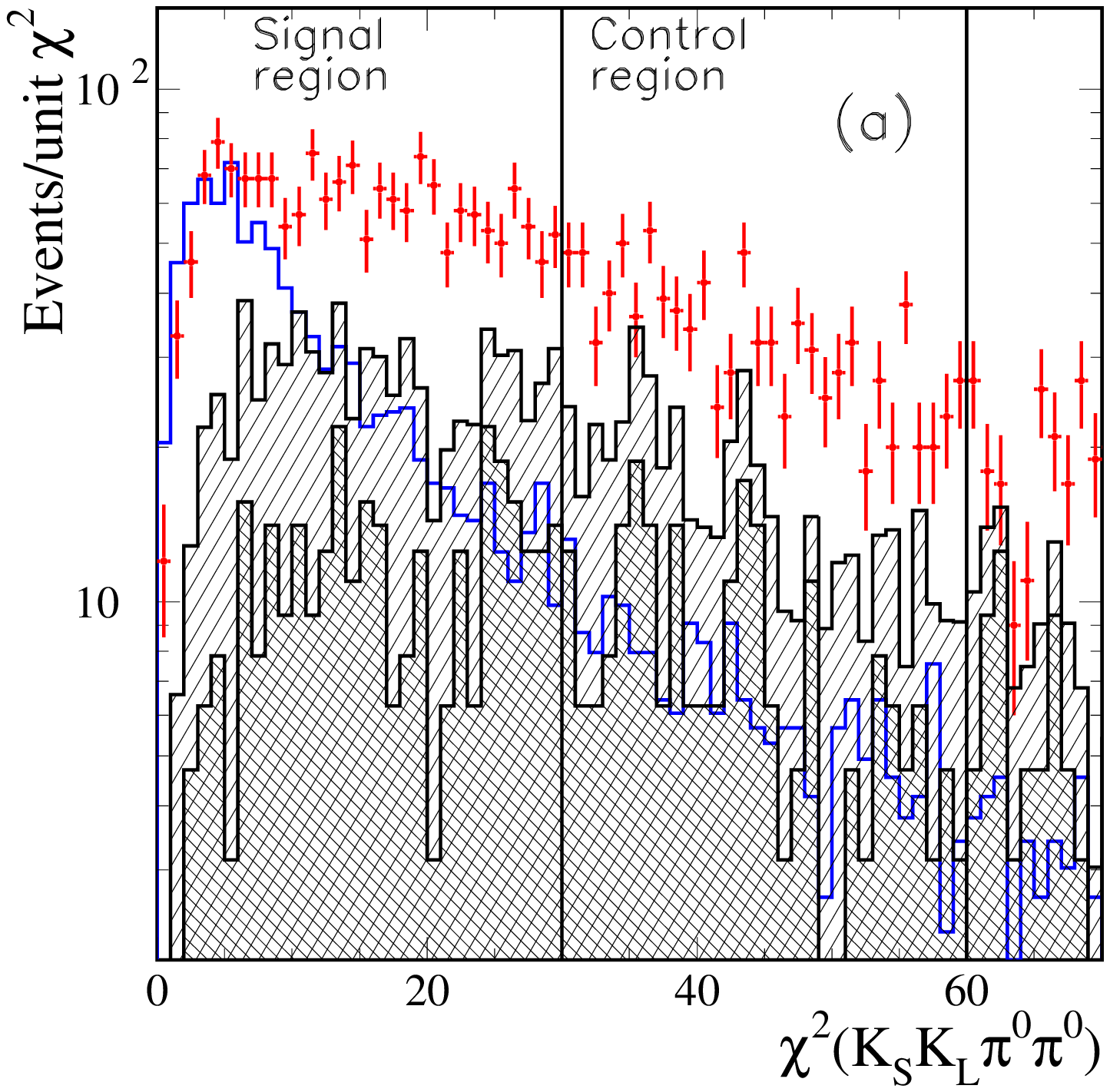}
\includegraphics[width=0.33\linewidth]{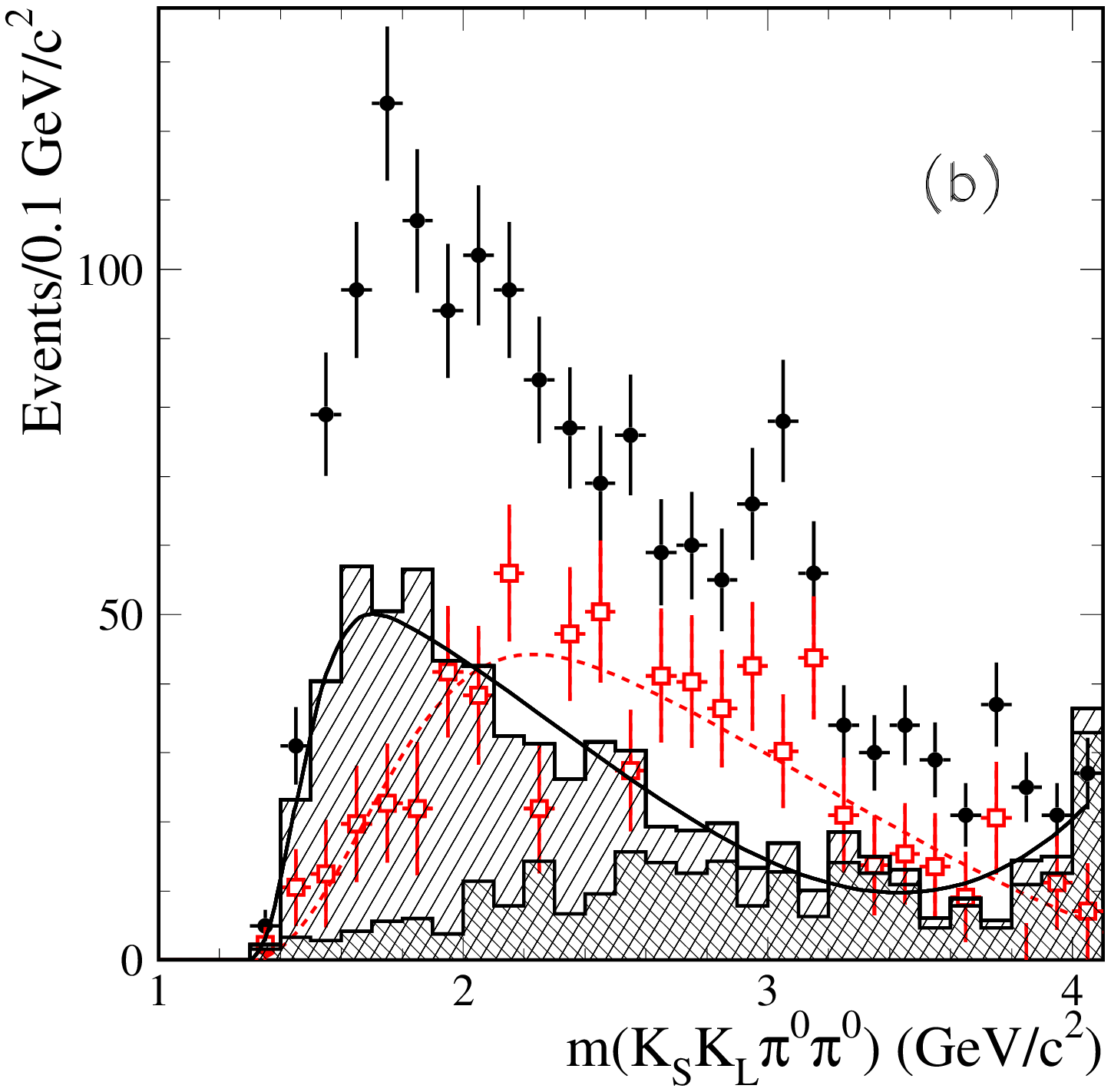}
\includegraphics[width=0.33\linewidth]{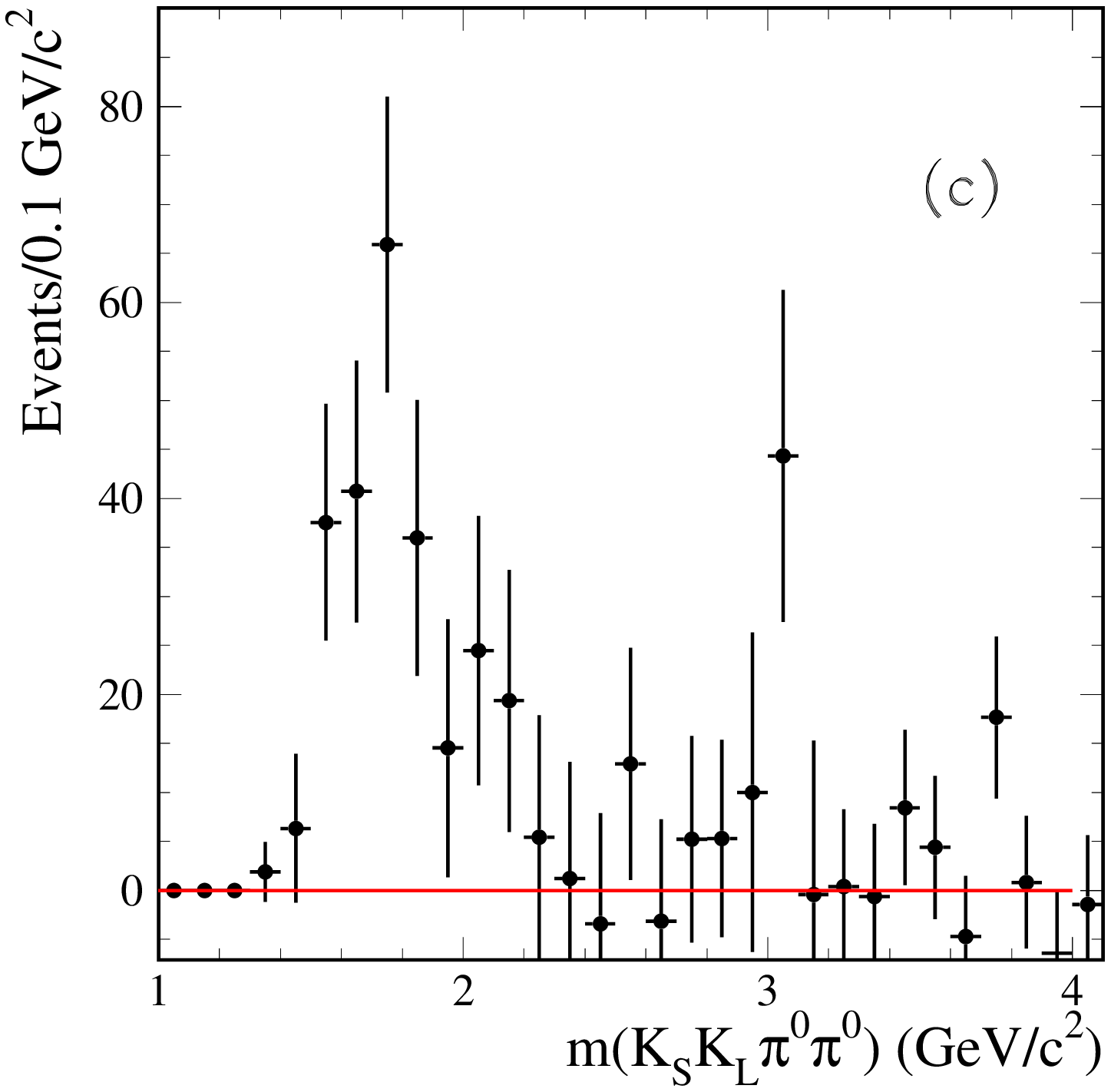}
\vspace{-0.5cm}
\caption{
  (a) The five-constraint \chisq distributions for data (points) and
      MC-simulated $\KKppz\gamma$ events (open histogram).
      The cross-hatched and hatched areas represent the
      simulated backgrounds from non-ISR $q\bar q$, and the sum of ISR $\KS\KL$,
      \KKpz,  and \KKeta events, respectively. 
  (b) The \KKppz invariant mass distribution for data events in the
      signal region of (a) (points).
      The cross-hatched, and hatched areas represent the simulated
      contributions from non-ISR $q\bar q$ events, and the sum of
      known ISR events, respectively,
      and the open squares represent the additional background
      estimated from the control region. 
      The curves show the empirical fits used for background subtraction.
  (c) The \KKppz invariant mass distribution after subtraction
      of all backgrounds.
}
\label{kskl2pi0_chi2}
\end{center}
\end{figure*}
\subsection{Final selection and backgrounds}
\label{addkskl2pi0}
From all events with a \KS, a \KL and at least two non-overlapping
\piz candidates, 
we consider the combination with the best value of \chiKKppg, 
as described in Sec.~\ref{sec:Analysis}.
Since background candidates are not well suppressed using additional photon or \piz,
no additional requirements are imposed.
Figure~\ref{kskl2pi0_chi2}(a) shows the \chiKKppg distribution of the
data (dots),
compared with that of the signal simulation (open histogram). 
The simulated distribution is normalized to the data in the region
$\chiKKppg<5$, 
where the contribution of higher-order ISR is small and the background
contamination is lowest,
but still amounts to about 20\% of the signal. 
The cross-hatched and hatched areas represent the simulated
backgrounds from non-ISR $q\bar q$ and 
the sum of ISR $\KS\KL$, \KKpz, and \KKeta events, respectively,
where the latter two are normalized to our measurements reported
above.
Together, these account for a substantial fraction of the entries at
high \chiKKppg values. 

We define a signal region $\chiKKppg<30$ 
and a control region $30<\chiKKppg<60$ 
(vertical lines in Fig.~\ref{kskl2pi0_chi2}(a)),
containing 1748 data and 2465 signal-MC signal,
and 990 data and 517 signal-MC events, respectively.
The $m(\KKppz)$ distribution for the events in the signal region
is shown in Fig.~\ref{kskl2pi0_chi2}(b) as the points. 
The cross-hatched and hatched areas show the simulated contributions from 
non-ISR $q\bar q$ events, and the sum of the ISR $\KS\KL$, \KKpz, and
\KKeta events, respectively.

We use a two-step procedure to subtract backgrounds in this final
state. 
From all experimental distributions,
we first subtract the normalized MC-simulated events just discussed.
In the case of the $m(\KKppz)$ distribution,
we fit an empirical function to the sum of these, 
shown as the solid line in Fig.~\ref{kskl2pi0_chi2}(b),
and use that for a bin-by-bin subtraction.
A similar procedure is applied to all other distributions,
including that of \chiKKppg (see Fig.~\ref{kskl2pi0_chi2}(a)).

We then use events from the \chiKKppg control region,
after subtraction of the backgrounds just described,
to calculate the remaining background in each bin of each distribution,
as described in Sec.~\ref{sec:ksklpi0bkg}.
We show this contribution to the $m(\KS\KL\ppz)$ distribution by the
open squares in Fig.~\ref{kskl2pi0_chi2}(b).
We fit a smooth function to reduce fluctuations,
and use the results (dotted curve in Fig.~\ref{kskl2pi0_chi2}(b)) to
subtract the remaining background.

After subtraction of all backgrounds,
we obtain 392$\pm$55 signal events with masses between threshold and
4.0~\gevcc, 
distributed as shown in Fig.~\ref{kskl2pi0_chi2}(c).  
We estimate the systematic uncertainty due to backgrounds
to be about 25\% of the signal for $m(\KKppz)<2.2$~\gevcc, 
increasing roughly linearly to 100\% at 3.0~\gevcc, 
and everywhere smaller than the statistical uncertainty.
There is no significant signal above about 2~\gevcc,
apart from an indication of the \jpsi and \psitwos signals in the
0.1~\gevcc wide bins.

\begin{figure}[tbh]
\begin{center}
\includegraphics[width=0.9\linewidth]{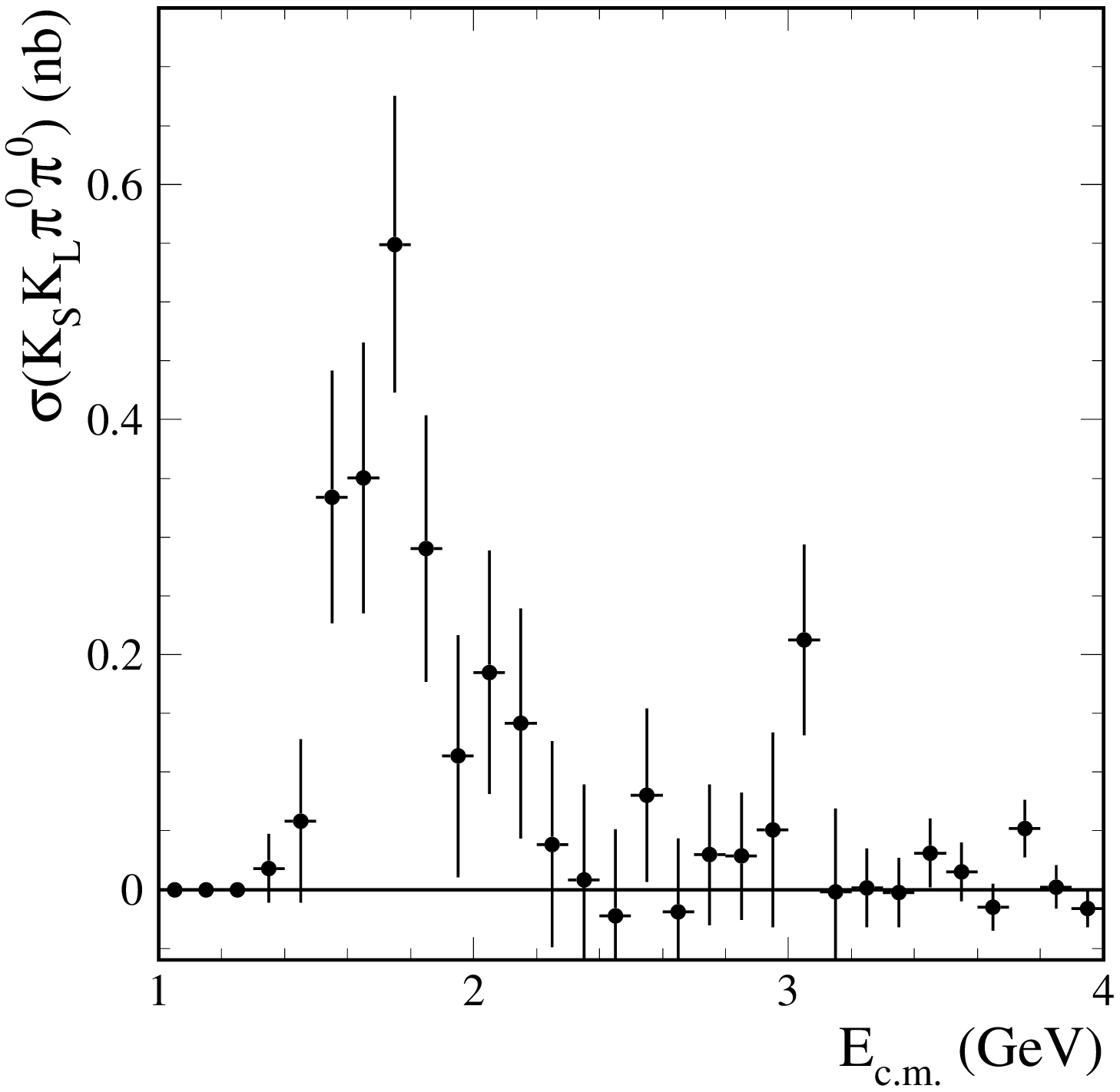}
\vspace{-0.4cm}
\caption{
The $\epem\to \KKppz$ cross section.
}
\label{kskl2pi0_xs}
\end{center}
\end{figure}
\subsection{\boldmath Cross section for $\epem\to \KKppz$ }
\label{sec:xskskl2pi0}
We calculate the $\epem \!\!\to\!\KKppz$  cross section as a
function of the effective c.m.\ energy \Ecm using
Eq.~\ref{xsformular}.  
The simulated efficiency is 1.5\% and shows no dependence on the 
\KKppz invariant mass. 
All corrections discussed in Sec.~\ref{sec:xsksklpi0} are applied,
plus an additional 3\% for the detection of the second \piz. 
The fully corrected cross section is shown in Fig.~\ref{kskl2pi0_xs}
and listed in Table~\ref{kskl2pi0_tab},
with statistical uncertainties only.
There are no other measurements for this final state. 

The cross section shows a rise from a threshold at 1.4~\gev,
a maximum value of about 0.5 nb near 1.8~\gev,
and a decrease with increasing energy.
Apart from a \jpsi and possibly a \psitwos signal (discussed below), the
cross section is statistically consistent with zero above 2.2~\gev.
The relative systematic uncertainty is dominated by the uncertainty of the
backgrounds, 
and totals 25\% at the peak of the cross section, 
increasing linearly to about 60\% at 2~\gev, and 100\% at higher energies.
\begin{table}
\caption{Summary of the $\epem\to \KS\KL\ppz$ 
cross section measurement. Uncertainties are statistical only.}
\label{kskl2pi0_tab}
\begin{ruledtabular}
\begin{tabular}{ c c c c }
$E_{\rm c.m.}$ (GeV) & $\sigma$ (nb)  
& $E_{\rm c.m.}$ (GeV) & $\sigma$ (nb) 

\\
\hline

    1.3500 &  0.018 $\pm$  0.029 & 2.7500 &  0.029 $\pm$  0.059\\
    1.4500 &  0.059 $\pm$  0.070 & 2.8500 &  0.028 $\pm$  0.054\\
    1.5500 &  0.334 $\pm$  0.108 & 2.9500 &  0.051 $\pm$  0.083\\
    1.6500 &  0.350 $\pm$  0.115 & 3.0500 &  0.213 $\pm$  0.081\\
    1.7500 &  0.549 $\pm$  0.126 & 3.1500 & -0.002 $\pm$  0.071\\
    1.8500 &  0.290 $\pm$  0.113 & 3.2500 &  0.002 $\pm$  0.033\\
    1.9500 &  0.113 $\pm$  0.103 & 3.3500 & -0.002 $\pm$  0.029\\
    2.0500 &  0.185 $\pm$  0.104 & 3.4500 &  0.031 $\pm$  0.029\\
    2.1500 &  0.141 $\pm$  0.098 & 3.5500 &  0.015 $\pm$  0.025\\
    2.2500 &  0.038 $\pm$  0.088 & 3.6500 & -0.015 $\pm$  0.020\\
    2.3500 &  0.008 $\pm$  0.081 & 3.7500 &  0.052 $\pm$  0.024\\
    2.4500 & -0.022 $\pm$  0.073 & 3.8500 &  0.002 $\pm$  0.018\\
    2.5500 &  0.080 $\pm$  0.074 & 3.9500 & -0.016 $\pm$  0.016\\
    2.6500 & -0.019 $\pm$  0.062 &        &                    \\

\end{tabular}
\end{ruledtabular}
\end{table}

\begin{figure}[tbh]
\begin{center}
\includegraphics[width=0.98\linewidth]{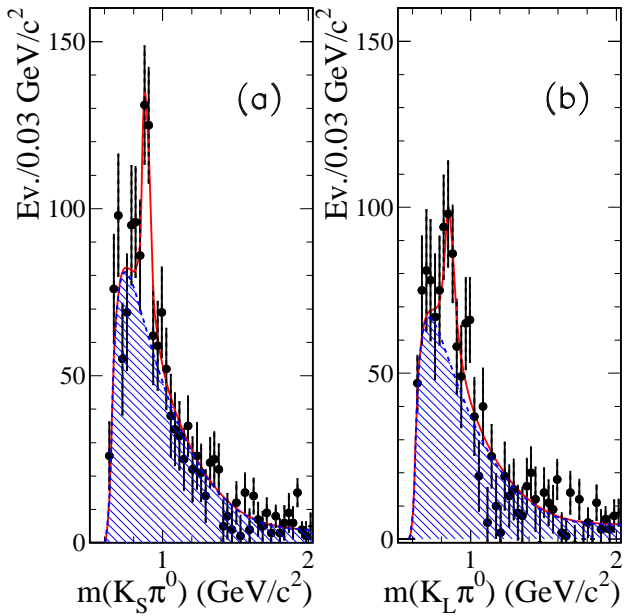}
\vspace{-0.4cm}
\caption{
The (a) $\KS\piz$ and (b) $\KL\piz$ invariant mass distributions for
selected \KKppz events in the data (points). 
The curves represent the results of the fits described in the text,
with the hatched areas representing the non-resonant components.
}
\label{kstarpi2}
\end{center}
\vspace{-0.4cm}
\end{figure}
\subsection{\boldmath The $K^{*}(892)^0$ and $\phi$ contributions } 
\label{sec:kstar2}
Figure~\ref{kstarpi2} shows the $\KS\piz$ and the $\KL\piz$ invariant
mass distributions for the selected \KKppz events after background
subtraction (two entries per event).
Signals corresponding to the $K^{*}(892)^0$ resonance are evident,
but the statistics are not sufficient to study them in detail.
\begin{figure*}[th]
\begin{center}
\includegraphics[width=0.33\linewidth]{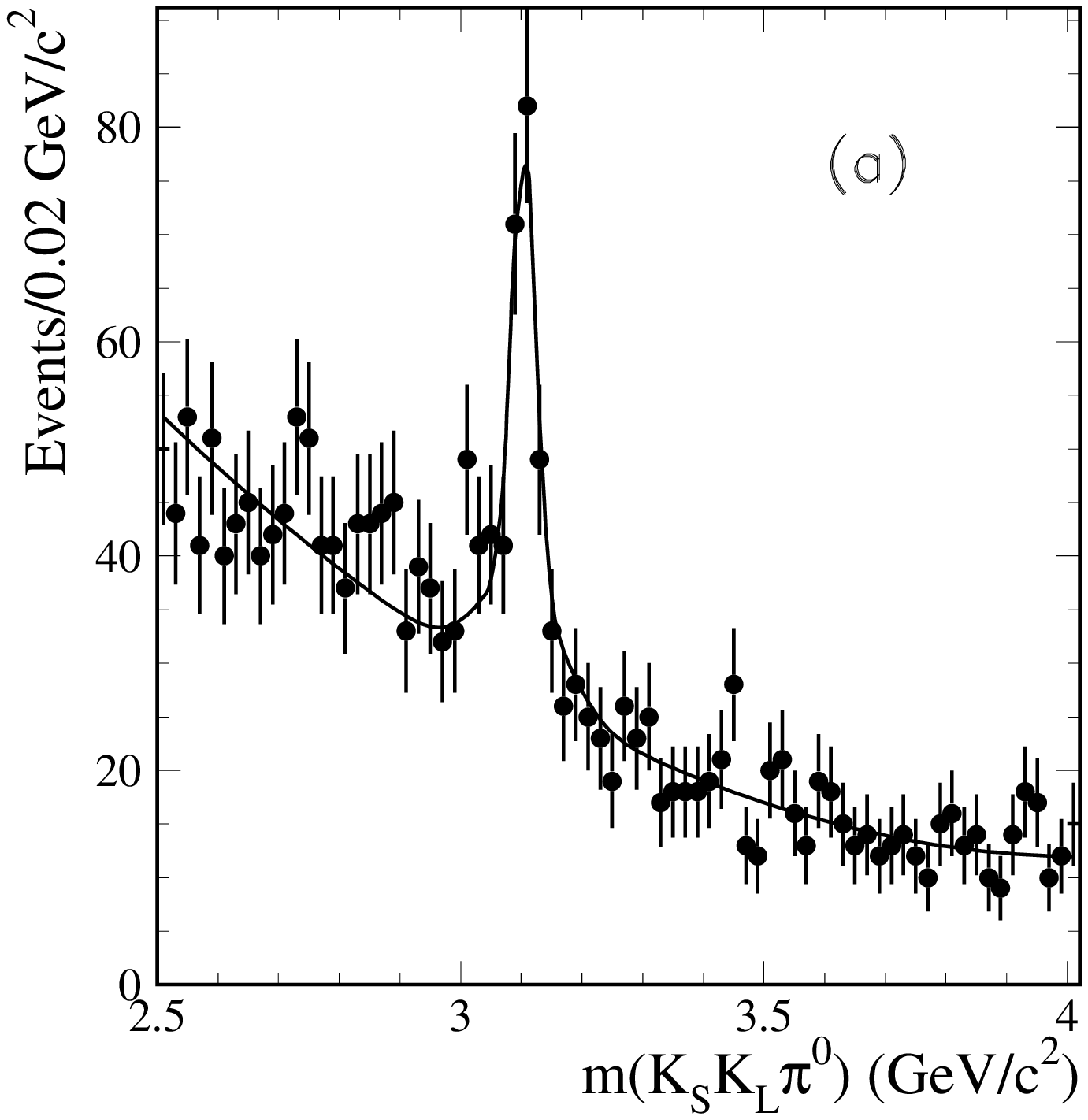}
\includegraphics[width=0.33\linewidth]{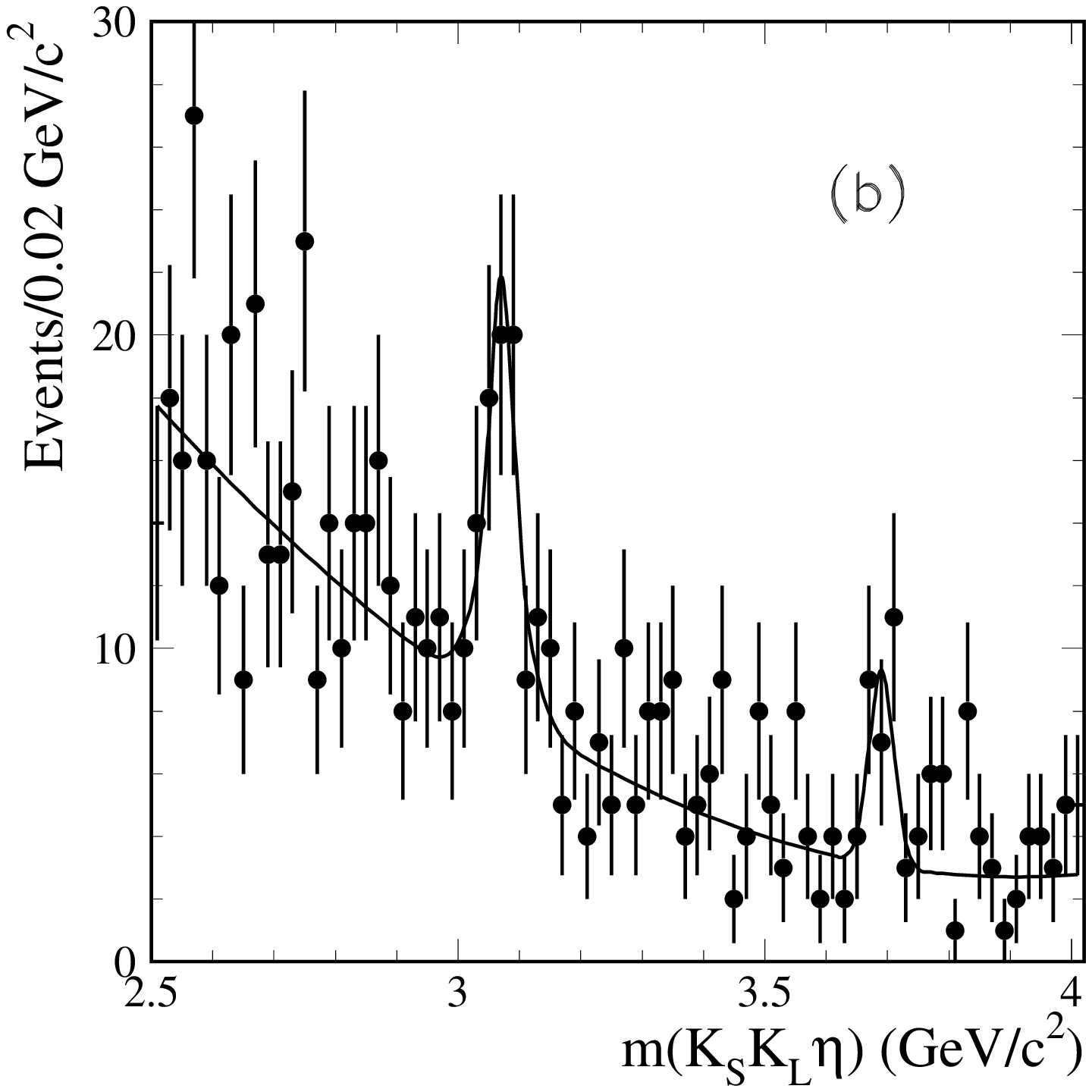}
\includegraphics[width=0.33\linewidth]{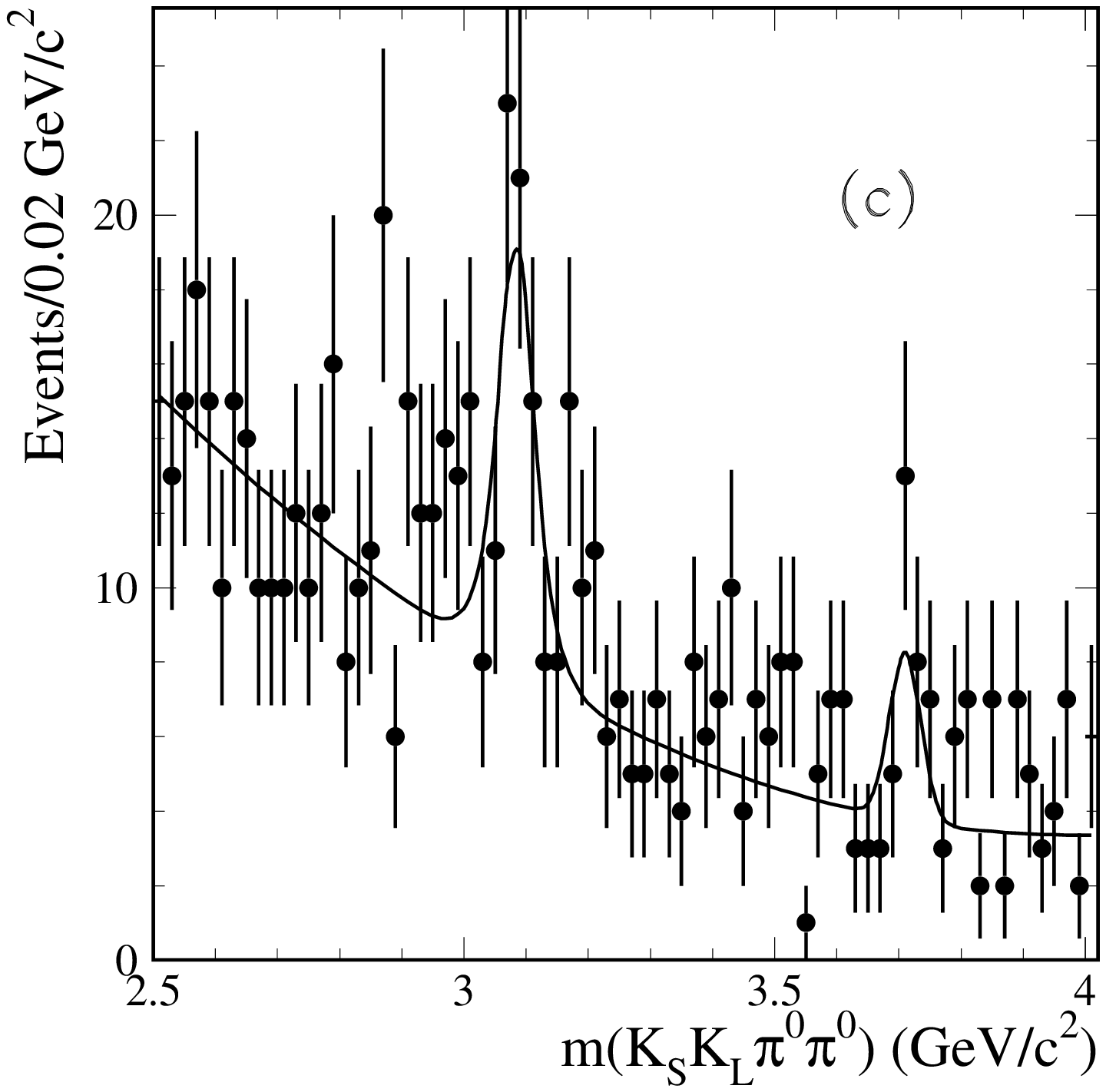}
\vspace{-0.4cm}
\caption{
Expanded views of the invariant mass distributions in the charmonium
mass region for the (a) $\KS\KL\piz$,  (b) $\KS\KL\eta$, 
and (c) $\KS\KL\ppz$ final states.
The lines represent the results of the fits described in the text.
}
\label{kskljpsi}
\end{center}
\end{figure*}

As an exercise,
we fit these distributions with a sum of a Breit-Wigner function
and a smooth function describing the non-resonant contribution,
yielding $190\pm44$ $K^{*}(892)^0\to \KS\piz$ decays, and
$171\pm32$ $K^{*}(892)^0\to \KL\piz$ decays.
There is no indication from the scatter plot (not shown) of 
any contribution from the $\epem\to K^{*}(892)^0 \Kbar^*(892)^0$ reaction,
and the sum of the two $K^{*}(892)$ yields is less than the total
number of \KKppz events,
limiting any such contribution to half the signal events.
This pattern is consistent with the dominance of the 
$\epem\to K^{*}(892)^0 \Km\pip + c.c.$ processes
seen in our previous measurement of the $\epem\to\Kp\Km\pip\pim$
reaction~\cite{isr2k2pi}.

Fitting the $\KS\KL$ invariant mass distribution (not shown),
we observe $71\pm16$ events from the $\epem\to\phi(1020)\ppz\to\KKppz$
process.
This is consistent with expectations from our measurement in the
$\Kp\Km\ppz$ mode~\cite{isr2k2pi}, 
but with substantially lower precision.
\section{The charmonium region}
\label{sec:jpsi}
Figure~\ref{kskljpsi} shows expanded views of the mass
distributions in Figs.~\ref{ksklpi0_chi2}(b), \ref{kskleta_chi2}(b),
and~\ref{kskl2pi0_chi2}(b), respectively,
in the 2.5--4.0~\gevcc mass region without any background subtraction. 
There are clear \jpsi signals in all three distributions, and
indications of \psitwos signals in (b) and (c).
Fitting the distribution in Fig.~\ref{kskljpsi}(a) with the sum of the
simulated \jpsi signal shape and a second-order polynomial function
yields $182\pm21$ $\jpsi\to\KKpz$ decays. 
No signal from the $\psitwos\to\KKpz$ decay
is observed ( $<$ 8 events at 90\% C.L.).
Fitting the other two distributions with the sum of simulated \jpsi and
\psitwos signal shapes and a second-order polynomial function yields 
$ 45\pm10$ $\jpsi\to\KKeta$ decays,
$ 47\pm11$ $\jpsi\to\KKppz$ decays, 
$ 16\pm 5$ $\psitwos\to\KKeta$ decays, and 
$ 14\pm 6$ $\psitwos\to\KKppz$ decays.

Using the corrected simulated efficiencies described above and the
differential luminosity,
we calculate the products of the \jpsi ($\psi(2S)$) electronic width and
branching fractions to these modes, and list them in Table~\ref{jpsitab}.
Using the PDG value of 
$\Gamma_{ee}(\jpsi) = 5.55$~\kev 
($\Gamma_{ee}(\psi(2S)) = 2.35$~\kev)~\cite{PDG},  
we obtain the corresponding branching fractions, 
also presented in Table~\ref{jpsitab}. 
Systematic uncertainties of typically 5\% arise from the corrections
to the simulated efficiencies, discussed in Sec.~\ref{sec:ksklpi0eff}, 
and variations of the signal shape.

These are the first observations of these three \jpsi decay modes.
Our \KKpz branching fraction 
can be compared with existing measurements of similar modes
$\BR(\jpsi\to\Kp\Km\piz)=(2.8\pm0.8)\times 10^{-3}$~\cite{franklin}
and $\BR(\jpsi\to\Kpm\KS\pimp)=(2.6\pm0.7)\times10^{-3}$~\cite{vannucci}.
The data are consistent with the expectation from isospin conservation
that they are equal.
Our other two measured branching fractions are consistent with
existing results for the corresponding modes involving charged
kaons~\cite{PDG},
of $\BR(\jpsi\to \Kp\Km\eta) = (0.85\pm0.14)\times 10^{-3}$ and
$\BR(\jpsi\to \Kp\Km\ppz) = (2.35\pm0.41)\times 10^{-3}$, respectively.

There are no previous observations of \psitwos decays into any of
these modes.
Our measurements indicate the presence of the $\psitwos\to\KKeta$
and \KKppz decay modes at just over three and two standard
deviations, respectively,  and we give in Table ~\ref{jpsitab} an upper
limit at the 90\% C.L on the \KKpz mode.
\begin{table*}[tbh]
\caption{
  Summary of the \jpsi and \psitwos   
  branching fractions
  obtained in this analysis.
  }
\label{jpsitab}

\newcolumntype{d}[1]{D{.}{.}{#1}}

\begin{ruledtabular}
\begin{tabular}{r@{~$\cdot$}l            @{\hspace*{.8cm}} 
                d{5.3}@{$\pm$}l@{$\pm$}l @{\hspace*{.8cm}}
                d{5.3}@{$\pm$}l@{$\,\pm$}l @{\hspace*{1.5cm}} cl } 
\multicolumn{2}{c}{Measured} & \multicolumn{3}{c}{Measured \hspace*{0.3cm} }  &
\multicolumn{5}{c}{Calculated Branching Fractions (10$^{-3}$)}\\
\multicolumn{2}{c}{Quantity} & \multicolumn{3}{c}{Value (\ev)\hspace*{0.3cm} } &
\multicolumn{3}{l}{\hspace*{0.8cm} This work}    & 
\multicolumn{2}{l}{\hspace*{-0.3cm} Previous} \\
\hline
$\Gamma^{\jpsi}_{ee}$   &  $\BR_{\jpsi  \to \KS\KL\piz}$        &
 11.4  & 1.3  & 0.6   &  2.06 & 0.24 & 0.10  &  --  &                      \\

$\Gamma^{\jpsi}_{ee}$   &  $\BR_{\jpsi  \to \KS\KL\eta}$        &
  8.0  & 1.8  & 0.4   &  1.45 & 0.32 & 0.08  & -- &                      \\

$\Gamma^{\jpsi}_{ee}$   &  $\BR_{\jpsi  \to \KS\KL\ppz} $  &
 10.3  & 2.3  & 0.5   &  1.86 & 0.43 & 0.10  &  -- &                      \\

$\Gamma^{\jpsi}_{ee}$   &  $\BR_{\jpsi  \to K^*(892)^0 \Kbar^0 + c.c.} 
                          \!\cdot\!   \BR_{K^*(892)^0\to K^0\piz}$  &
  6.7  & 0.9  & 0.4   &  1.20 & 0.15 & 0.06  &  -- &                      \\

$\Gamma^{\jpsi}_{ee}$   &  $\BR_{\jpsi  \to K_2^*(1430)^0\Kbar^0 + c.c.}  
                          \!\cdot\!   \BR_{K_2^{*}(1430)\to K^0\piz}$  &
  2.4  & 0.7  & 0.1   &  0.43 & 0.12 & 0.02  & $<4$ & \cite{PDG} \\

$\Gamma^{\psitwos}_{ee}$ &  $\BR_{\psitwos \to \KS\KL\piz}$ &
\multicolumn{3}{l}{\hspace*{0.6cm} $<0.7$} & 
\multicolumn{3}{l}{\hspace*{0.4cm} $<0.3$} & -- & \\

$\Gamma^{\psitwos}_{ee}$ &  $\BR_{\psitwos \to \KS\KL\eta}$ &
  3.14 & 1.08~ & 0.16   &  1.33 & 0.46 & 0.07 & -- &\\

$\Gamma^{\psitwos}_{ee}$ &  $\BR_{\psitwos \to \KS\KL\ppz}$ &
  2.92 & 1.27~ & 0.15   &  1.24 & 0.54 & 0.06 & -- &\\

\end{tabular}
\end{ruledtabular}
\end{table*}
\begin{figure}[tbh]
\begin{center}
\vspace{-0.2cm}
\includegraphics[width=1.08\linewidth]{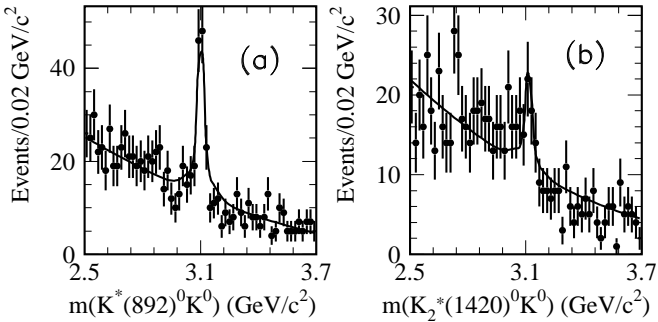}
\vspace{-0.4cm}
\caption{ 
The $\KS\KL\piz$ invariant mass distributions for  events
in which the $\KS\piz$ or $\KL\piz$ mass is within 0.15~\gevcc of
(a) the nominal $K^*(892)^0$ mass or (b) the nominal $K^*_2(1430)^0$ mass.
The lines represent the results of the fits described in the text.
}
\label{jpsikstark}
\end{center}
\end{figure}
\subsection{\boldmath Internal structure of the
$\jpsi\to\KKpz$, \KKeta, and \KKppz decays}

We consider \KKpz events in the charmonium region with a
$\KS\piz$ or $\KL\piz$ invariant mass within 0.15~\gevcc of the
nominal $K^*(892)^0$ or $K^*_2(1430)^0$ mass, 
and show their \KKpz invariant mass distributions in
Figs.~\ref{jpsikstark}(a) and \ref{jpsikstark}(b),
respectively. 
Fits using simulated \jpsi signal shapes and polynomial backgrounds
yield $106\pm13$ $\jpsi\to(K^*(892)^0 \Kbar^0 + c.c.)\to \KKpz$
events
and $37\pm11$ $\jpsi\to (K^*_2(1430)^0 \Kbar^0 + c.c.)\to \KKpz$
events.
For each of these intermediate states we calculate the product of its
\jpsi branching fraction, $\Gamma^{J/\psi}_{ee}$, and the relevant
branching fractions for the intermediate resonances,
and list the values in Table~\ref{jpsitab}.
Using $\Gamma^{J/\psi}_{ee} = 5.55$~\ev~\cite{PDG}
we calculate the corresponding products of branching fractions.

This first measurement of 
$\BR(\jpsi\to (K^*_2(1430)^0\Kbar^0+ c.c.)\to K^0\Kbar^0\piz)$
is consistent with the existing upper limit of $4\times10^{-3}$~\cite{PDG}.
According to isospin relations,
the $\jpsi\to(K^*(892)^0 \Kbar^0 + c.c.)\to\KKpz$ decay rate
should be the same as the existing world average of
$1.97\pm0.20\times10^{-3}$~\cite{PDG} for the charged-kaon decay chain 
$\jpsi\to(K^*(892)^+ \Kbar^- + c.c.)\to\Kp\Km\piz$,
and a factor of two lower than the $3.2\pm0.4\times10^{-3}$~\cite{PDG}
observed in the $\Kbar^0\Kp\pim + c.c.$ final state.
Our result is consistent with the latter expectation, 
and 2.5 standard deviations below the former.
In Sec.~\ref{sec:xsphieta} we noted that the $\phi\eta$ contribution to the
\jpsi signal in the \KKeta mode is very small.
We estimate $5\pm3$ events, corresponding to
$\BR(\jpsi\to\phi\eta)=(0.52\pm0.32)\times 10^{-3}$,
which is consistent with the PDG~\cite{PDG} value of
$(0.75\pm0.08)\times 10^{-3}$.  
The \jpsi signal in the \KKppz mode has high background 
(see Fig.~\ref{kskljpsi}(c)),
and we are unable to quantify the contributions from the 
$K^*(892)^0 K^0\piz$ and $\phi\ppz$ intermediate states with reasonable
accuracy.
The $\jpsi\to\phi\ppz$ decay rate is relatively well
measured~\cite{PDG}, dominated by our previous measurement in the
$\Kp\Km\ppz$ final state.

\section{Summary}
We have presented studies of the processes
$\epem\to\KKpz$, $\epem\to \KKeta$, and $\epem\to \KKppz$ 
at center-of-mass energies below 4~\gev, 
using events with initial-state radiation collected with the
\babar\ detector.
The cross sections for all three processes are measured for the first
time, over the energy range from threshold to 4~\gev, 
and their resonant structure is studied.

The $\epem\to \KKpz$ cross section is measured with 10--30\%
systematic uncertainty below 3~\gev, 
and is similar in shape to the $\epem\to\Kp\Km\piz$ cross section~\cite{isrkkpi}.
It is dominated by resonant, quasi-two body intermediate states.
The $K^{*}(892)^0\Kbar^0+ c.c.$ processes account for about 90\% of
the cross section, 
and there are few-percent contributions from the
$K^{*}(1430)^0\Kbar^0+ c.c.$ and $\phi\piz$ processes.
The cross section for the latter is consistent with that measured
previously in the $\Kp\Km\piz$ final state.

The $\epem\to \KKeta$ cross section is measured with 15--30\%
systematic uncertainty below 3~\gev, 
and is similar to the $\epem\to\Kp\Km\eta$ cross section~\cite{isrkkpi}.
The $\phi\eta$ intermediate state dominates below 2.0~\gev and
contributes up to 3.0~\gev, 
and its cross section is consistent with that measured previously in the
$\Kp\Km\eta$ final state.
No other intermediate states are observed, 
and non-resonant $\KKeta$ production is substantial in the 2.2--3.0~\gev range.

The $\epem\to\KKppz$ cross section is measured with 25--60\%
systematic uncertainty below 3~\gev,
and is consistent with the $\epem\to\KS\KS\pip\pim$ cross
section~\cite{isrkskl}.
Its \Ecm behavior is similar in shape to the $\epem\to\Kp\Km\pip\pim$,
$\KS\KL\pip\pim$, and $\Kp\Km\ppz$ cross sections, but factors of
about 8, 2, and 1.5 smaller, respectively.
There are substantial, but not dominant, contributions from
the $K^*(892)^0\Kbar^0\piz$, $K^{*}(1430)^0\Kbar^0\piz$ and
$\phi(1020)\ppz$ intermediate states.
These are hard to quantify, but are consistent with
expectations from our previous studies of the $\epem\to\Kp\Km\pip\pim$
and $\Kp\Km\ppz$ processes.
We see no evidence for any $K^{*0}\Kbar^{*0}$ intermediate states,
also consistent with the low rates we have observed in final states
involving charged kaons.


We observe the $\jpsi\to \KS\KL\piz$, $\KS\KL\eta$, and
$\KS\KL\ppz$ decays for the first time, 
and measure the product of the \jpsi electronic width and branching
fraction to each of these modes.
We study the resonant structure of these decays,
and obtain measurements of  
the $\jpsi\to K^*(892)^0 \Kbar^0 + c.c.$ and
$\jpsi\to K_2^*(1430)^0 \Kbar^0 + c.c.$ branching fractions times
$\Gamma^\jpsi_{ee}$.
In addition, 
we observe the $\psitwos\to\KKeta$ and $\psitwos\to\KKppz$ decays
for the first time, 
and measure the products of the \psitwos electronic width and the
corresponding branching fractions.

\section*{Acknowledgements} 
We are grateful for the 
extraordinary contributions of our \pep2\ colleagues in
achieving the excellent luminosity and machine conditions
that have made this work possible.
The success of this project also relies critically on the 
expertise and dedication of the computing organizations that 
support \babar.
The collaborating institutions wish to thank 
SLAC for its support and the kind hospitality extended to them. 
This work is supported by the
US Department of Energy
and National Science Foundation, the
Natural Sciences and Engineering Research Council (Canada),
the Commissariat \`a l'Energie Atomique and
Institut National de Physique Nucl\'eaire et de Physique des Particules
(France), the
Bundesministerium f\"ur Bildung und Forschung and
Deutsche Forschungsgemeinschaft
(Germany), the
Istituto Nazionale di Fisica Nucleare (Italy),
the Foundation for Fundamental Research on Matter (The Netherlands),
the Research Council of Norway, the
Ministry of Education and Science of the Russian Federation, 
Ministerio de Econom\'{\i}a y Competitividad (Spain), the
Science and Technology Facilities Council (United Kingdom),
and the Binational Science Foundation (U.S.-Israel).
Individuals have received support from 
the Marie-Curie IEF program (European Union) and the A. P. Sloan Foundation (USA). 


\end{document}